\documentclass[12pt]{article}
\usepackage[T1]{fontenc}
\usepackage[utf8]{inputenc}
\usepackage[english]{babel}
\usepackage{lmodern}
\usepackage{microtype}
\usepackage{authblk}
\usepackage{graphicx}
\usepackage{subcaption}
\usepackage{amssymb}
\usepackage{float}
\usepackage{amssymb}
\usepackage{amsmath}
\usepackage{wrapfig}
\usepackage{xcolor}
\usepackage[left=20mm, top=15mm, right=15mm, bottom=15mm, nohead, footskip=10mm]{geometry}

\title{Lorentz-Invariant Measure for Path Integrals in the Future Cone}

\author[a]{Vladimir V. Belokurov\thanks{vvbelokurov@yandex.ru}} 
\author[a]{Vsevolod V. Chistiakov\thanks{vsevolod.chistyakov@gmail.com}}
\author[a]{Klavdiia A. Lursmanashvili\thanks{klavo4ka2001@gmail.com}}
\author[a]{Evgeniy T. Shavgulidze\thanks{shavgulidze@bk.ru}}

\affil[a]{Lomonosov Moscow State University}

\date{}

\begin{document}
\maketitle
\abstract{We construct a measure on continuous paths in the future cone of two-dimensional space that is invariant under the orthochronous Lorentz group and quasi-invariant under the action of the diffeomorphism group.

We use the proposed measure to obtain the fundamental solution of the heat equation. The result agrees with the expression obtained by other methods. }

\section{Introduction}
 The construction of functional measures in quantum field theory and gravity is problematic because of the pseudo-Euclidean structure of spacetime. For example, quantum theories of vector fields, string theory \cite{Kaku}, and calculations within the worldline formalism \cite{Schubert} deal with path integrals of the form
\begin{equation}\label{rel PI}
    \int F(\xi) \exp\left(-\frac{1}{2}\int_0^T d\tau \,\,\left[(\dot \xi^0)^2-(\dot \xi^1)^2\right]\right)d\xi\, 
\end{equation}
(for simplicity, we consider the two-dimensional case).

The integrals (\ref{rel PI}) formally do not exist because of the factor $\sim\exp\left(+\frac{1}{2}\int d\tau\, (\dot \xi^1)^2\right)$ in the integrand. The standard procedure is to consider a family of integrals of the form
\begin{equation}\label{alphaint}
      \int F(\xi) \exp\left(-\frac{1}{2}\int_0^T d\tau \,\,\left[(\dot \xi^0)^2+\alpha(\dot \xi^1)^2\right]\right)d\xi\, ,
\end{equation}
depending on a complex parameter $\alpha$. For positive real $\alpha$, integral (\ref{alphaint}) is a Wiener integral. The value of integral (\ref{rel PI}) is then postulated to be the analytic continuation to $\alpha= -1$ of integral (\ref{alphaint}), initially evaluated for $\alpha>0$. The validity of this procedure is not evident, since it changes the symmetry of the integration measure. Integral (\ref{rel PI}) is invariant under the Lorentz group $SO^{\uparrow}(1,1)$, whereas the regularized integral (\ref{alphaint}) for $\alpha>0$ is invariant under the rotation group $SO(2)$.

Other attempts to define integral (\ref{rel PI}) \cite{Padm,Dudley} encounter similar difficulties.

In this paper, we propose a different approach to this problem. We construct a measure $\tilde w_{\sigma}$ on the space $C([0,T],\mathrm{Cone})$ of paths lying in the future cone
\begin{equation}\label{cone}
\mathrm{Cone}=\{(x^0,x^1);\,\, x^0>0,\,\, |x^1|<x^0\}\, ,    
\end{equation}
that is quasi-invariant under the diffeomorphism group and invariant under the group of orthochronous Lorentz transformations $SO^{\uparrow}(1,1)$.
The measure $\tilde w_{\sigma}$ has the form
\begin{equation}\label{result_measure}
    \tilde w_{\sigma}(d\xi)=\rho \,\dot \varphi(0)\dot \varphi(T)\mu_{\frac{2\sigma}{\rho}}(d\varphi)\,d\rho\, w_{\frac{\sigma}{\rho}}(d\psi)\, ,
\end{equation}
where
\begin{equation}
    \xi \in C([0,T],\mathrm{Cone})\, ,\qquad \rho \in \mathbb{R}^{+}\, ,\qquad \varphi \in \mathrm{Diff}_{+}^1([0,T])\, ,\qquad \psi\in C([0,T],\mathbb{R})\, , 
\end{equation}
$d\rho$ is the Lebesgue measure on the positive half-line, $\mu_{\frac{2\sigma}{\rho}}$ is the quasi-invariant measure on the diffeomorphism group, and $w_{\frac{\sigma}{\rho}}(d\psi)$ is the Wiener measure.

The construction of measure (\ref{result_measure}) uses the decomposition of the Wiener measure over the orbits of the diffeomorphism group (the orbital decomposition) \cite{polar,polar2dRJ}, described in {\bf Sections \ref{GeneralFacts} and \ref{Decomposition}}.

In {\bf Section \ref{OurMeasure}}, we construct measure (\ref{result_measure}) and formulate the {\bf main theorem}, which states that $\tilde w_{\sigma}$ corresponds to a stochastic process with independent increments (a Markov process). We then show that (\ref{result_measure}) is equivalent to the Wiener measure on the universal cover of the punctured plane $\mathbb{R}^2\backslash\{0\}$.

In {\bf Section \ref{Fundamental solution}}, we use the proposed measure to obtain the fundamental solution of the heat equation on $\mathrm{Cone}$ and show that the result agrees with expressions obtained by other methods in \cite{Smits,Ford,DowQFTCone}. We also establish a relation between the fundamental solutions of the heat equation on $\mathrm{Cone}$ and on $\mathbb{R}^2$.

In {\bf Section \ref{Conclusion}}, we discuss the results and their possible applications to other problems.

In {\bf Appendix A}, we express the measure $\tilde w_{\sigma}$ in terms of the Cartesian coordinates $\xi(\tau)=(\xi^0(\tau),\xi^1(\tau))$ of the stochastic process in $\mathrm{Cone}$. {\bf Appendix B} contains the proof of the {\bf main theorem}. In {\bf Appendix C} examples of paths in $\mathrm{Cone}$ and the corresponding paths on the universal cover of the punctured plane are given.

\section{Basic definitions and properties of the Wiener measure}\label{GeneralFacts}

On the space of continuous functions $C([t_0,t_0+T],\mathbb{R})$, the Wiener measure with parameter $\sigma$ is given by
\begin{equation}\label{win1d}
    w_{\sigma}(d\xi)=\exp\left(-\frac{1}{2\sigma^2}\int_{t_0}^{t_0+T} \dot \xi^2(\tau) d\tau\right)d\xi\, .
\end{equation}
Here $\xi\in C([t_0,t_0+T],\mathbb{R})$, and the initial value $\xi(t_0)$ is arbitrary.

The Wiener measure has the following basic properties.

\noindent Invariance under translations of the time variable\\
    \begin{equation}\label{shift}
         w_{\sigma}(d\xi)=w_{\sigma}(d\eta),\quad  \mbox{where} \quad  \eta(\tau)=\xi(\tau-t_1)\in C([t_0+t_1,t_0+t_1+T],\mathbb{R});
    \end{equation}
Transformations under rescaling of the time variable are given by the following equation\\
    \begin{equation}\label{scale}
         w_{\sigma}(d\xi)=w_{\sigma\sqrt{a}}(d\eta),\quad  \mbox{where} \quad  \eta(\tau_1)=\xi(a\tau_1)\in C\left(\left[\frac{t_0}{a},\frac{t_0+T}{a}\right],\mathbb{R}\right);
    \end{equation}
Transformations under rescaling of the function take the form\\
    \begin{equation}\label{scale1}
          w_{\sigma}(d\xi)=\lambda w_{\sigma/\lambda}(d\eta)   \qquad   \xi(\tau)=\lambda\eta(\tau)\, . 
    \end{equation}

These properties allow us to take $t_0=0$ and $T=1$ without loss of generality. In the main discussion, we use functions defined on the interval $[0,T]$, passing to the interval $[0,1]$ whenever convenient.

A standard result in the theory of stochastic processes \cite{Shyr1} states that, for any $t\in[0,T]$, the values of the Wiener process $\xi(\tau)$ on $[t,T]$ depend on its history on $[0,t)$ only through the value $\xi^{*}=\xi(t)$.

We split the process $\xi(\tau)$ on $[0,T]$ into two processes, $\xi_1(\tau)$ and $\xi_2(\tau)$, defined on $[0,t]$ and $[t,T]$, respectively:
\begin{equation}\label{concat}
    \xi=\Xi(\xi_1,\xi_2)\equiv \left\{\begin{array}{cc}
        \xi_1(\tau) &  \tau\in[0,t]\\
        \xi_2(\tau) &  \tau\in [t,T]
    \end{array}\right.
\end{equation}

For an arbitrary functional $F(\xi)$, the following identity holds:
\begin{equation}\label{mark}
    \int F(\xi)w_{\sigma}(d\xi)=\int \tilde  F(\xi_1,\xi_2)\delta(\xi_1(t)-\xi_2(t))w^1_{\sigma}(d\xi_1)w^2_{\sigma}(d\xi_2)\, ,
\end{equation}
where $w^{1}_{\sigma}$ is the Wiener measure on $C([0,t],\mathbb{R})$, $w^{2}_{\sigma}$ is the Wiener measure on $C([t,T],\mathbb{R})$, and
\begin{equation}
    \tilde F(\xi_1,\xi_2)\equiv F(\Xi(\xi_1,\xi_2))\, .
\end{equation}

Property (\ref{mark}) implies a semigroup property for an associated family of operators. Consider the function
\begin{equation}
     K(\tau,x,y)=\int\delta(\xi(0)-y)\delta(\xi(\tau)-x) w_{\sigma}(d\xi)
\end{equation}
From (\ref{mark}), it follows that this function $K$ satisfies the integral relation
\begin{equation}\label{conv}
     K(\tau_1+\tau_2,x,y)=\int_{\mathbb{R}} dz\,\,  K(\tau_2,x,z) K(\tau_1,z,y)\, .
\end{equation}
and the initial condition
\begin{equation}\label{IC}
    \lim_{\tau\to 0} K(\tau,x,y)=\delta(x-y)
\end{equation}

Equations (\ref{conv}) and (\ref{IC}) imply that the family of operators $\mathcal{U}_{\tau}$, defined by
\begin{equation}\label{U_t}
    (\mathcal{U}_\tau v)(x)=\int_{\mathbb{R}} dy\,\,  K(\tau,x,y) v(y)
\end{equation}
forms a semigroup:
\begin{equation}\label{semigroup}
\mathcal{U}_{\tau_2}\mathcal{U}_{\tau_1}=\mathcal{U}_{\tau_1+\tau_2}\, , \qquad \mathcal{U}_0=I\, .
\end{equation}
Here $I$ is the identity operator.

The function $K(\tau,x,y)$ (also known as the heat kernel) is the fundamental solution of the heat equation
\begin{equation}\label{heatfirst}
    \frac{\partial K}{\partial \tau}=\frac{\sigma^2}{2}\frac{\partial^2}{\partial x^2} K
\end{equation}
with the initial condition (\ref{IC}).

The preceding discussion extends naturally to spaces of vector-valued continuous functions. In particular, on the space $C([t_0,t_0+T],\mathbb{R}^2)$ of two-component continuous functions $\xi(\tau)=(\xi^0(\tau),\xi^1(\tau))$, the Wiener measure is defined by
\begin{equation}\label{win2d}
    w_{\sigma}(d\xi)=w_{\sigma}(d\xi^0)w_{\sigma}(d\xi^1)=\exp\left(-\frac{1}{2\sigma^2}\int_{t_0}^{t_0+T}d\tau\left[({\dot \xi^0}(\tau))^2+({\dot \xi^1}(\tau))^2\right]\right)d\xi^0d\xi^1\, .
\end{equation}
Properties (\ref{shift}) and (\ref{scale}) remain valid, whereas the property (\ref{scale1}) takes the form
\begin{equation}
    w_{\sigma}(d\xi)=\lambda^2 w_{\sigma/\lambda}(d\eta)\, ,\qquad \xi(\tau)=\lambda\eta(\tau)\, .
\end{equation}

The rotation group $SO(2)$ acts naturally on $\mathbb{R}^2$:
\begin{equation}\label{rot}
    R_{\gamma}\in SO(2)\, ,\qquad R_{\gamma}\xi(\tau)\equiv\left(\xi^0(\tau)\cos\gamma-\xi^1(\tau) \sin\gamma,\xi^0(\tau)\sin\gamma+\xi^1(\tau)\cos\gamma \right)\, .
\end{equation}
Measure (\ref{win2d}) is invariant under (\ref{rot}):
\begin{equation}\label{Rinv}
    w_{\sigma}(d(R_{\gamma}\xi))=w_{\sigma}(d\xi)\, .
\end{equation}
In two dimensions, the analogue of equation (\ref{heatfirst}) is
\begin{equation}\label{heat2d}
    \frac{\partial K}{\partial \tau}=\frac{\sigma^2}{2}\Delta K\, ,\qquad \mbox{where}\quad  \Delta=\frac{\partial^2}{\partial (x^0)^2}+\frac{\partial^2}{\partial (x^1)^2}\, .
\end{equation}
The solution of (\ref{heat2d}) in polar coordinates
\begin{equation}
    x=(r\cos\theta,r\sin\theta)\, ,\qquad y=(r_0\cos\theta_0,r_0\sin\theta_0)\, ,
\end{equation}
has the form
\begin{equation}\label{2dkern}
    K(\tau,x,y)=\frac{1}{2\pi\tau\sigma^2}\exp\left(-\frac{r^2+r_0^2-2rr_0\cos(\theta-\theta_0)}{2\tau\sigma^2}\right)\, .
\end{equation}

{\bf Quasi-invariance under the diffeomorphism group.}\\
Let $\varphi\in \mathrm{Diff}^1_{+}([0,T])$. The diffeomorphism group acts on the space $C([0,T],\mathbb{R})$ as follows:\footnote{Here $\dot \varphi^{-1}(\tau)\equiv\frac{d}{d\tau}\varphi^{-1}(\tau)$.}
\begin{equation}\label{diffact}
    (\varphi \xi)(\tau)=\frac{\xi(\varphi^{-1}(\tau))}{\sqrt{\dot\varphi^{-1}(\tau)}}
\end{equation}
The Wiener measure is quasi-invariant under the action of the subgroup $\mathrm{Diff}_{+}^3([0,T])$ \cite{Shepp,Shavg}:
\begin{equation}
    w_{\sigma}(d(\phi\xi))=\mathcal{P}_{\phi}(\xi)w_{\sigma}(d\xi)\,, \qquad \phi\in\mathrm{Diff}_{+}^3([0,T]) .
\end{equation}
Here
\begin{equation}
    \mathcal{P}_{\varphi}(\xi)=\sqrt[4]{\dot\varphi(0)\dot \varphi(T)}\exp\left(\frac{1}{4\sigma^2}\left[(\xi(0))^2\frac{\ddot \varphi(0)}{\dot \varphi(0)}-(\xi(T))^2\frac{\ddot \varphi(T)}{\dot \varphi(T)}\right]\right)\exp\left(\frac{1}{4\sigma^2}\int_0^Td\tau \,\,(\xi(\tau))^2\mathrm{Sch}\{\varphi,\tau\}\right)\, 
\end{equation}
is the Radon--Nikodym derivative, and
\begin{equation}
    \mathrm{Sch}\{\varphi,\tau\}\equiv\frac{d}{d\tau}\left(\frac{\ddot \varphi}{\dot \varphi}\right)-\frac12\left(\frac{\ddot \varphi}{\dot \varphi}\right)^2\, 
\end{equation}
is the Schwarzian derivative.

The measure (\ref{win2d}) is quasi-invariant under $\mathrm{Diff}^3_{+}([0,T])$, just as the Wiener measure on $C([0,T],\mathbb{R})$ is. The action of $\mathrm{Diff}_{+}^1([0,T])$ on $C([0,T],\mathbb{R}^2)$ is defined as follows. In addition to the Cartesian coordinates $\xi^0$ and $\xi^1$, we introduce the polar coordinates $r(\tau)$ and $\theta(\tau)$:
\begin{equation}
    \xi^0(\tau)=r(\tau)\cos\theta(\tau)\,, \qquad \xi^1(\tau)=r(\tau)\sin\theta(\tau)\, .
\end{equation}
The group $\mathrm{Diff}_{+}^1([0,T])$ acts as
\begin{equation}\label{act2diff}
    (\varphi \xi)(\tau)=\left(\frac{r(\varphi^{-1}(\tau))}{\sqrt{\dot \varphi^{-1}(\tau)}}\cos\theta(\varphi^{-1}(\tau)),\frac{r(\varphi^{-1}(\tau))}{\sqrt{\dot \varphi^{-1}(\tau)}}\sin\theta(\varphi^{-1}(\tau))\right)\, .
\end{equation}
The action (\ref{act2diff}) of $\mathrm{Diff}_{+}^1([0,T])$ commutes with the action (\ref{rot}) of the rotation group: $R_{\gamma}\varphi \xi=\varphi R_{\gamma}\xi$.

\section{Orbital decomposition of the Wiener measure under $\mathrm{Diff}_{+}^1([0,T])$}\label{Decomposition}

Consider the space $C_{+}([0,T],\mathbb{R})\subset C([0,T],\mathbb{R})$ of continuous positive functions on $[0,T]$. This space decomposes into orbits of the action (\ref{diffact}) of $\mathrm{Diff}^1_{+}([0,T])$ \cite{polar}. The integral
\begin{equation}
    \frac{1}{\rho^{2}}\equiv\frac{1}{T}\int_0^T \frac{d\tau}{\xi^2(\tau)}
\end{equation}
is invariant under the transformation (\ref{diffact}).

Using the orbital decomposition of $C_{+}([0,T],\mathbb{R})$ under $\mathrm{Diff}_{+}^1([0,T])$, we represent a function $\xi(\tau)\in C_{+}([0,T],\mathbb{R})$ as follows. On each orbit, choose the constant function $\tilde \xi_{\rho}(\tau)=\rho$. All other elements of the orbit with fixed $\rho$ are obtained by applying the diffeomorphism group action (\ref{diffact}) to $\tilde \xi_{\rho}$. 

Thus, every element $\xi(\tau)\in C_{+}([0,T],\mathbb{R})$ can be represented as \cite{polar}
\begin{equation}
    \xi(\tau)=(\varphi\tilde\xi_{\rho})(\tau)=\rho\frac{1}{\sqrt{\dot\varphi^{-1}(\tau)}}\, .
\end{equation}
Each element $\xi\in C_{+}([0,T],\mathbb{R})$ uniquely determines a pair $(\rho,\varphi)$ of the decomposition coordinates, with $\rho\in \mathbb{R}^{+}$ and $\varphi\in \mathrm{Diff}^1_{+}([0,T])$, according to
\begin{equation}
    \frac{1}{\rho^2}=\frac{1}{T}\int_0^T\frac{d\tau_1}{\xi^2(\tau_1)}\, ,\qquad \varphi^{-1}(\tau)=\rho^2\int_0^{\tau}\frac{d\tau_1}{\xi^2(\tau_1)}\, .
\end{equation}

The restriction of the Wiener measure (\ref{win1d}) to the subspace $C_{+}([0,T],\mathbb{R})\subset C([0,T],\mathbb{R})$ admits the following decomposition over the orbits of $\mathrm{Diff}^1_{+}([0,T])$ \cite{polar}:
\begin{equation}\label{decomp1d}
    w_{\sigma}(d\xi)=\exp\left(-\frac{\sigma^2T}{8\rho^2}\right)(\dot \varphi(0)\dot \varphi(T))^{3/4}\mu_{\frac{2\sigma}{\rho}}(d\varphi)d\rho\, ,
\end{equation}
where $\mu$ is the quasi-invariant measure on the diffeomorphism group \cite{unusual view}, \cite{calc}:
\begin{equation}\label{mu}
    \mu_{\sigma}(d\varphi)=\frac{1}{\sqrt{\dot \varphi(0)\dot \varphi(T)}}\exp\left\{\frac{1}{\sigma^2}\left[\frac{\ddot \varphi(0)}{\dot \varphi(0)}-\frac{\ddot \varphi(T)}{\dot \varphi(T)}\right]\right\}\exp\left\{\frac{1}{\sigma^2}\int_0^T\mathrm{Sch}\{\varphi,\tau\}\, \, d\tau\right\}d\varphi\, .
\end{equation}
Here, $d\varphi$ denotes the generalized Haar measure on $\mathrm{Diff}_{+}^1([0,T])$, symbolically written as
\begin{equation}
    d\varphi\equiv\prod_{\tau_i\in[0,T]}\frac{d\varphi(\tau_i)}{\dot \varphi(\tau_i)}\, .
\end{equation}

The invariant of the diffeomorphism group action (\ref{act2diff}) on $C([0,T],\mathbb{R}^2)$ is
\begin{equation}
\frac{1}{\rho^{2}}=\frac{1}{T}\int_0^T  \frac{d\tau}{(\xi^0(\tau))^2+(\xi^1(\tau))^2}\, .
\end{equation}

We choose on each orbit of $\mathrm{Diff}^1_{+}([0,T])$ the representative of the form \cite{polar2d}
\begin{equation}\label{one element}
    \tilde \xi_{\rho,\psi,\alpha}(\tau)=\Big(\rho\cos\left[\psi(\tau)+\alpha\right],\rho\sin\left[\psi(\tau)+\alpha\right]\Big)\, ,
\end{equation}
where $\psi(\tau)\in C_0([0,T],\mathbb{R})$ and $\alpha\in S^1$. Here, $C_{0}([t_0,t_0+T],\mathbb{R})$ is the space of continuous functions with the fixed initial value $\xi(t_0)=0$, and $S^1=\mathbb{R}/2\pi\mathbb{Z}$ is the circle.

An arbitrary function $\xi\in C([0,T],\mathbb{R}^2)$ can be obtained from a function of the form (\ref{one element}) by the transformation (\ref{act2diff}):
\begin{equation}
    \xi=(\varphi\tilde \xi_{\rho,\psi,\alpha})(\tau)=\left(\frac{\rho}{\sqrt{\dot \varphi^{-1}(\tau)}}\cos\left[\psi(\varphi^{-1}(\tau))+\alpha\right]\, ,\frac{\rho}{\sqrt{\dot\varphi^{-1}(\tau)}}\sin\left[\psi(\varphi^{-1}(\tau))+\alpha\right]\,\right) .
\end{equation}
Thus, there is a one-to-one correspondence between continuous functions $\xi(\tau)\in C([0,T],\mathbb{R}^2)$ and four decomposition coordinates $(\rho,\psi,\varphi,\alpha)$, where $\rho\in \mathbb{R}^{+}$, $\psi\in C_0([0,T],\mathbb{R})$, $\varphi\in \mathrm{Diff}^1_{+}([0,T])$, and $\alpha\in S^1$. We denote this correspondence by
\begin{equation}
    \xi\simeq (\rho,\psi,\varphi,\alpha)\, . 
\end{equation}

\begin{equation}
    \frac{1}{\rho^2}=\frac{1}{T}\int_0^T\frac{d\tau_1}{(\xi^0(\tau_1))^2+(\xi^1(\tau_1))^2}\, ,\qquad \varphi^{-1}(\tau)=\rho^2\int_0^{\tau}\frac{d\tau_1}{(\xi^0(\tau_1))^2+(\xi^1(\tau_1))^2}
\end{equation}
\begin{equation}
    z(\tau)=\xi^0(\tau)+i\xi^1(\tau)\, ,\qquad \psi(s)=\mathrm{arg} \,z(\varphi(s))-\mathrm{arg}\,z(0)\, , \qquad \alpha=\mathrm{Arg}\, z(0)\, ,
\end{equation}
where
$$z(\tau)=\xi^0(\tau)+i\xi^1(\tau)\, .$$

It is readily seen that
\begin{equation}\label{gamma decomp}
    R_{\gamma}\xi\simeq(\rho,\psi,\varphi,\alpha+\gamma)\, .
\end{equation}

The decomposition of the Wiener measure over the orbits of the action (\ref{act2diff}) of $\mathrm{Diff}_{+}^1([0,T])$ takes the form \cite{polar2d}
\begin{equation}\label{decomp2d}
    w_{\sigma}(d\xi)=\rho\dot \varphi(0)\dot \varphi(T) \mu_{\frac{2\sigma}{\rho}}(d\varphi)w^{0}_{\frac{\sigma}{\rho}}(d\psi)d\rho d\alpha\, .
\end{equation}
Here, $w^0_{\sigma}$ is the Wiener measure on $C_0([0,T],\mathbb{R})$. 
Property (\ref{Rinv}) follows directly from (\ref{gamma decomp}) and (\ref{decomp2d}).

\section{A measure on paths in $\mathrm{Cone}$}\label{OurMeasure}
\subsection{A quasi-invariant measure on $C([0,T],\mathrm{Cone})$}
Now we construct a measure on $C([0,T],\mathrm{Cone})$. The two-dimensional space  $\mathrm{Cone}$ is defined by (\ref{cone}).
By analogy with the action of $SO(2)$ on the two-dimensional Euclidean plane (\ref{act2diff}), consider the action of $SO^{\uparrow}(1,1)$ (the Lorentz group) on $\mathrm{Cone}$:
\begin{equation}\label{Lorentz}
    L_{\gamma}\xi(\tau)=(\,\xi^0(\tau)\,\cosh \gamma + \xi^1(\tau)\,\sinh\gamma ,\xi^0(\tau)\,\sinh\gamma + \xi^1(\tau)\,\cosh\gamma)\, .
\end{equation}

Define the pseudopolar coordinates $r,\theta$ by
\begin{equation}
    \xi^0=r(\tau)\cosh(\theta(\tau))\, ,\qquad \xi^1=r(\tau)\sinh(\theta(\tau))\, .
\end{equation}
The action of the diffeomorphism group on $C([0,T],\mathrm{Cone})$ is given by
\begin{equation}\label{pseudo act}
    (\varphi \xi)(\tau)=\left(\frac{r(\varphi^{-1}(\tau))}{\sqrt{\dot \varphi^{-1}(\tau)}}\cosh\theta(\varphi^{-1}(\tau)),\frac{r(\varphi^{-1}(\tau))}{\sqrt{\dot \varphi^{-1}(\tau)}}\sinh\theta(\varphi^{-1}(\tau))\right)\, .
\end{equation}
The action commutes with the Lorentz transformations (\ref{Lorentz}).

The integral
\begin{equation}
    \frac{1}{\rho^{2}}=\frac{1}{T}\int_0^T\frac{d\tau}{(\xi^0(\tau))^2-(\xi^1(\tau))^2}
\end{equation}
is invariant under the transformations (\ref{pseudo act}).
On each orbit of $\mathrm{Diff}_{+}^1([0,T])$, choose an element of the form
\begin{equation}\label{one element pseudo}
    \tilde \xi_{\rho,\psi}=(\rho\cosh \psi(\tau),\rho\sinh\psi(\tau))\, ,
\end{equation}
where $\psi\in C([0,T],\mathbb{R})$. Every element $\xi\in C([0,T],\mathrm{Cone})$ can be obtained from an element of the form (\ref{one element pseudo}) by the transformation (\ref{pseudo act}):
\begin{equation}
    \xi(\tau)=(\varphi\tilde \xi_{\rho,\psi})(\tau)=\left(\frac{\rho}{\sqrt{\dot\varphi^{-1}(\tau)}}\cosh\psi(\varphi^{-1}(\tau))\, , \frac{\rho}{\sqrt{\dot \varphi^{-1}(\tau)}}\sinh\psi(\varphi^{-1}(\tau))\right)\, .
\end{equation}

Thus, each element $\xi \in C([0,T],\mathrm{Cone})$ corresponds to the set of three variables $(\rho,\psi,\varphi)$, where $\rho>0, \psi(\tau)\in C([0,T],\mathbb{R})$, $\varphi\in \mathrm{Diff}^1_{+}([0,T])$. We write this correspondence as
\begin{equation}
    \xi\simeq(\rho,\psi,\varphi)\, .
\end{equation}
\begin{equation}\label{decompcone}
    \frac{1}{\rho^2}=\frac{1}{T}\int_0^T\frac{d\tau_1}{(\xi^0(\tau_1))^2-(\xi^1(\tau_1))^2}\, ,\qquad \varphi^{-1}(\tau)=\rho^2\int_0^{\tau}\frac{d\tau_1}{(\xi^0(\tau_1))^2-(\xi^1(\tau_1))^2}\, , 
\end{equation}
\begin{equation*}
    \psi(s)=\mathrm{arctanh}\left(\frac{\xi^1(\varphi(s))}{\xi^0(\varphi(s))}\right)\, .
\end{equation*}
It is readily verified that
\begin{equation}\label{Lorentz_inv}
    L_{\gamma}\xi\simeq(\rho,\psi+\gamma,\varphi)\, .
\end{equation}

By analogy with the Wiener measure on $C([0,T],\mathbb{R}^2)$, we define the measure on $C([0,T],\mathrm{Cone})$ through its decomposition over the orbits of the diffeomorphism group:
\begin{equation}\label{wtilde}
     \tilde w_{\sigma}(d\xi)=\rho\dot \varphi(0)\dot \varphi(T) \mu_{\frac{2\sigma}{\rho}}(d\varphi)w_{\frac{\sigma}{\rho}}(d\psi)d\rho\, .
\end{equation}
From (\ref{Lorentz_inv}), (\ref{wtilde}) it follows that $\tilde w_{\sigma}$ is invariant under the transformations (\ref{Lorentz}).

\subsection{The metric on the cone and its properties}

Changing variables in (\ref{wtilde}) from $(\rho,\psi,\varphi)$ to the original variables $\xi^0(\tau)$, $\xi^1(\tau)$, we write the measure $\tilde w_{\sigma}$ in the form (see Appendix A)
\begin{equation}\label{tildewxi}
    \tilde w_{\sigma}(d\xi)=\exp\left(-\frac{1}{2\sigma^2}\int_0^T \langle \dot \xi,\dot \xi\rangle \,\,d\tau\right)d\xi\,  ,
\end{equation}
where
\begin{equation}\label{Conemetric}
    \langle \dot \xi,\dot \xi\rangle\equiv(\dot \xi^0)^2-(\dot \xi^1)^2+2\frac{(\xi^0\dot \xi^1-\xi^1\dot \xi^0)^2}{(\xi^0)^2-(\xi^1)^2}\, .
\end{equation}

The quadratic form (\ref{Conemetric}) defines a metric on the cone. We now describe its properties. The line element is\footnote{Here, coordinates on $\mathrm{Cone}$ are denoted by $(x^0,x^1)$. The letter $\xi$ denotes a stochastic process on $\mathrm{Cone}$.}
\begin{equation}\label{Conemet1}
    ds^2\equiv\langle dx,dx\rangle= (dx^0)^2-(dx^1)^2+2\frac{(x^0dx^1-x^1dx^0)^2}{(x^0)^2-(x^1)^2}\,=
\end{equation}
\begin{equation*}
    =\frac{(x^0)^2+(x^1)^2}{(x^0)^2-(x^1)^2}\left((dx^0)^2+(dx^1)^2\right)-4\frac{x^0x^1}{(x^0)^2-(x^1)^2}dx^0 dx^1\, =
\end{equation*}
\begin{equation*}
    =x^{+}x^{-}\left(\left(\frac{dx^{+}}{x^{+}}\right)^2+\left(\frac{dx^{-}}{x^{-}}\right)^2\right)=e^{l^{+}+l^{-}}((dl^{+})^2+(dl^{-})^2)\, ,
\end{equation*}
where
\begin{equation}
    x^{\pm}=\frac{x^0\pm x^1}{\sqrt{2}}\,  \qquad  l^{\pm}=\ln x^{\pm}\, .
\end{equation}

Introduce the pseudopolar coordinates $(r,\theta)$ by
\begin{equation}\label{x0x1hyp}
    x^0=r\cosh\theta\,, \qquad x^1=r\sinh\theta\, .
\end{equation}
The metric (\ref{Conemet1}) then takes the form of the Euclidean metric on $\mathbb{R}^2$:
\begin{equation}\label{polarmet}
    ds^2=dr^2+r^2d\theta^2\, .
\end{equation}
Note that in (\ref{x0x1hyp}) and (\ref{polarmet}), the ``angular'' coordinate $\theta$ can take any real value ($\theta\in\mathbb{R}$).

{\bf Remark}

In quantum gravity, there is the problem of performing the Wick rotation: the substitution $t\to-it_{E}$ generally leads to different Euclidean theories for different choices of the time coordinate $t$. To avoid this problem, complex-valued metrics were proposed in \cite{Konts}, \cite{Witten}. In this setting, the exponents in integrals of the form (\ref{alphaint}) contain expressions invariant under $SO(2,\mathbb{C})$. Note that both $SO(2,\mathbb{R})$ and $SO^{\uparrow}(1,1)$ are the subgroups (real forms) of $SO(2,\mathbb{C})$. In this sense, the measure $\tilde w_{\sigma}$ can be regarded as an analytic continuation of the Wiener measure $w_{\sigma}$.   

The measure $\tilde w_{\sigma}$ can be used to define integrals of the form (\ref{rel PI}). Since the metric (\ref{Conemet1}) takes the form (\ref{polarmet}) in pseudopolar coordinates, the Minkowski metric can be obtained by analytic continuation in the angular coordinate. Specifically, instead of (\ref{polarmet}), we consider, for $\alpha>0$, 
\begin{equation}
    ds^2=dr^2+\alpha \,r^2d\theta^2,
\end{equation}
and set $\alpha=-1$ after completing all calculations. This approach differs from analytic continuation in Cartesian coordinates (\ref{alphaint}).

\subsection{An isometry from the cone to the universal cover of the punctured plane}

The space $\mathrm{Cone}$, equipped with the metric (\ref{Conemet1}), is isometric to the universal (infinitely sheeted) cover of the punctured plane. Recall the standard construction of the universal cover of $\mathbb{R}^2\backslash\{0\}$. Consider an infinite collection of planes
\begin{equation}
    \Pi_n=\{(r,\theta); r>0,\theta\in [0,2\pi)\}\, ,
\end{equation}
where each point of $\Pi_n$ is specified by its polar coordinates $r,\theta$: the distance $r$ from the origin and the polar angle $\theta$. Each plane $\Pi_n$ is equipped with the metric (\ref{polarmet}). Cut every plane $\Pi_n$ along the direction $\theta=0$ and glue the planes along these cuts so that a point on $\Pi_n$, after making one complete counterclockwise revolution around the origin, arrives on $\Pi_{n+1}$. In other words, set
\begin{equation}
    \lim_{\theta\to2\pi} (r,\theta)_n=(r,0)_{n+1}\, .
\end{equation}
We denote the collection of planes $\Pi_n$ glued in this way by $\mathrm{Cover}$. Each point of $\mathrm{Cover}$ can be specified by a triple $(r,\theta,n)$, where $r,\theta$ are polar coordinates on one of the planes $\Pi_n$, with $\theta\in[0,2\pi)$, and $n$ is the plane index. Alternatively, a point can be specified by a pair $(r,\theta)$ with $\theta\in \mathbb{R}$, with the convention that for $2\pi n\le \theta<2\pi(n+1)$ the point belongs to $\Pi_n$ and its polar angle on $\Pi_n$ is $\theta-2\pi n$.

Thus, each point of $\mathrm{Cover}$ is specified by two coordinates, $r\in \mathbb{R}^{+}$ and $\theta\in \mathbb{R}$. The metric on $\mathrm{Cover}$ is inherited from the planes $\Pi_{n}$ and therefore has the form (\ref{polarmet}). It is readily seen that $\mathrm{Cone}$ and $\mathrm{Cover}$ are in one-to-one correspondence, obtained by identifying points with the same coordinates:
\begin{equation}\label{conecov}
    (r,\theta)\in \mathrm{Cone}\longleftrightarrow(r,\theta)\in\mathrm{Cover}\, .
\end{equation}
This correspondence is an isometry, since the metric (\ref{polarmet}) on $\mathrm{Cone}$ coincides with the metric (\ref{polarmet}) on $\mathrm{Cover}$. Under this mapping, each plane $\Pi_n$ corresponds to a sector of $\mathrm{Cone}$ (the sector between the red dashed lines in Fig.~1).

\begin{figure}[H]
    \centering
    \includegraphics[width=0.5\linewidth]{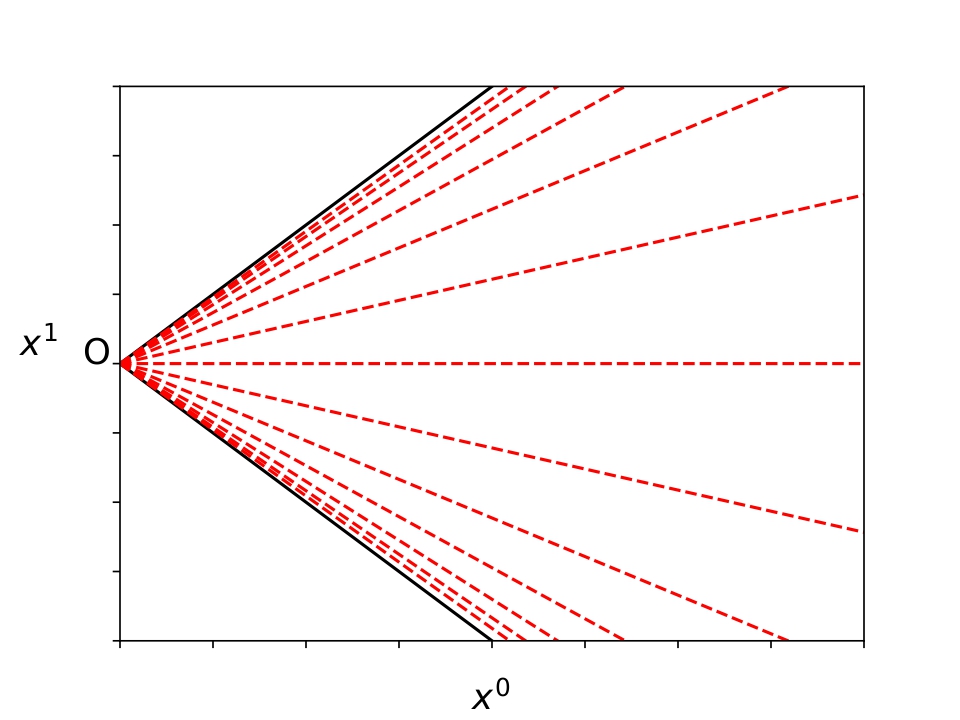}
    \caption{The sector between the red dashed lines is the preimage of a single sheet of the cover.}
    \label{fig:placeholder}
\end{figure}

\subsection{Main theorem}\label{Maintheorem}
We first state a result proved in \cite{polar2dRJ}, which implies property (\ref{mark}) for the measure given by (\ref{decomp1d}). Without loss of generality, assume that $\xi(\tau)$ is defined on $[0,1]$. As in (\ref{concat}), split the process $\xi(\tau)$ into two subprocesses $\xi_1(\tau)$ and $\xi_2(\tau)$ on $[0,t]$ and $[t,1]$, respectively. Rescaling the argument as in (\ref{scale}), replace $\xi_1(\tau), \xi_2(\tau)$ by $\eta_1(\tau),\eta_2(\tau)\in C_{+}([0,1],\mathbb{R})$. On the one hand, $\eta_1(\tau)$ and $\eta_2(\tau)$ are expressed in their decomposition coordinates as
\begin{equation}\label{own}
    (\eta_1)_{\varphi_1,\rho_1}(\tau)=\frac{\rho_1}{\sqrt{\dot \varphi_1^{-1}(\tau)}}\, ,\qquad (\eta_2)_{\varphi_2,\rho_2}(\tau)=\frac{\rho_2}{\sqrt{\dot \varphi^{-1}_2(\tau)}}\, .
\end{equation}
On the other hand, the chain of one-to-one correspondences $(\rho,\varphi)\simeq\xi\longleftrightarrow \eta_1,\eta_2$ allows us to express $\eta_1(\tau)$ and $\eta_2(\tau)$ in terms of the decomposition coordinates of the original process $\xi(\tau)$:
\begin{equation}\label{first_decomp}
    (\eta_{1})_{\rho,\varphi}(\tau)=\frac{\rho}{\sqrt{\dot \varphi^{-1}(t\tau)}}\, ,\quad (\eta_2)_{\rho,\varphi}(\tau)=\frac{\rho}{\sqrt{\dot \varphi^{-1}(t+(1-t)\tau)}}\, .
\end{equation}
The following lemma states the equality of the integrals in the parameterizations (\ref{own}) and (\ref{first_decomp}).\\
{\bf Lemma }\\
For every bounded Borel function $F: C_{+}([0,1],\mathbb{R})\times C_{+}([0,1],\mathbb{R}) \to \mathbb{R}$, $t\in (0,1)$, and $a>0$, the following identity holds:
\begin{equation}
    \int_0^{+\infty} \int_{\mathrm{Diff}_{+}^1([0,1])} F((\eta_1)_{\rho,\varphi},(\eta_2)_{\rho,\varphi})\exp\left(-\frac{a^2}{\sigma^2}\rho^2\dot \varphi(0)-\frac{\sigma^2}{8\rho^2}\right)(\dot \varphi(0)\dot \varphi(1))^{3/4}\mu_{\frac{2\sigma}{\rho}}(d\varphi)d\rho=
\end{equation}
\begin{equation*}
\int_{0}^{+\infty}\int_{0}^{+\infty}\int_{\mathrm{Diff}_{+}^1([0,1])}\int_{\mathrm{Diff}_{+}^1([0,1])}\delta\left(\rho_1\sqrt{\dot\varphi_1(1)}-\rho_2\sqrt{\dot \varphi_2(0)}\right)F((\eta_1)_{\varphi_1,\rho_1},(\eta_2)_{\varphi_2,\rho_2})\times
\end{equation*}
\begin{equation*}
    \exp\left(-\frac{a^2}{t\sigma^2}\rho_1^2\dot \varphi_1(0)-\frac{t\sigma^2}{8\rho_1^2}-\frac{(1-t)\sigma^2}{8\rho_2^2}\right)(\dot \varphi_1(0)\dot \varphi_1(1))^{3/4}(\dot \varphi_2(0)\dot \varphi_2(1))^{3/4}\mu_{\frac{2\sigma \sqrt{t}}{\rho_1}}(d\varphi_1)\mu_{\frac{2\sigma\sqrt{1-t}}{\rho_2}}(d\varphi_2)d\rho_1d\rho_2\, .
\end{equation*}

\vspace{1cm}

We now state a theorem that implies property (\ref{mark}) for the measure (\ref{wtilde}) constructed above. Its structure is analogous to that of the preceding lemma. Specifically, the process $\xi\in C([0,1],\mathrm{Cone})$ is split into two processes $\xi_1,\xi_2\in C([0,1],\mathrm{Cone})$. Both processes are then expressed in their own decomposition coordinates, $\xi_1\longleftrightarrow(\rho_1,\psi_1,\varphi_1)$ and $\xi_2\longleftrightarrow(\rho_2,\psi_2,\varphi_2)$, and in terms of the decomposition coordinates $\xi\longleftrightarrow(\rho,\psi,\varphi)$ of the original process. It is convenient to use pseudopolar coordinates:
\begin{equation}
    \xi_1(\tau)=r_1(\tau)(\cosh\theta_1(\tau),\sinh\theta_1(\tau))\, ,\qquad \xi_2(\tau)=r_2(\tau)(\cosh\theta_2(\tau),\sinh\theta_2(\tau))\, .
\end{equation}

For $0\leq \tau \leq 1$, set
$$(r_1)_{\varphi ,\rho} (\tau)=\frac{\rho}{\sqrt{\dot \varphi ^{-1}(t \tau )}}\,, \qquad (r_2)_{\varphi ,\rho} (\tau)=\frac{\rho}{\sqrt{\dot \varphi ^{-1}(t+(1-t) \tau )}}\,, $$
$$(r_1)_{\varphi _1,\rho_1} (\tau)=\frac{\rho_1}{\sqrt{\dot\varphi_1 ^{-1}(\tau )}}\, ,\qquad (r_2)_{\varphi_2,\rho_2} (\tau)=\frac{\rho_2}{\sqrt{\dot\varphi_2 ^{-1}(\tau )}}\, .$$

where
$$t \in (0,1)\, , \quad \varphi,\varphi_1\, ,\varphi_2\in \mathrm{Diff}^1_{+}([0,1]),\, \quad \rho>0 , \, \rho_1>0 , \, \rho_2 >0\, , $$
and suppose that
$$\rho_1 \sqrt{\dot \varphi _1 (1)}=\rho_2 \sqrt{\dot \varphi _2 (0)}\, .$$

Note that
$(r_1)_{\varphi ,\rho}, (r_2)_{\varphi ,\rho}, 
(r_1)_{\varphi _1,\rho_1}, (r_2)_{\varphi _2,\rho_2} \in C_+ ([0,1],\mathbb{R}).$
 
Let
$$\psi, \psi_1, \psi_2  \in C([0,1],\mathbb{R})\, . $$
Set
$$(\theta_1)_{\varphi,\psi}(\tau)=\psi(\varphi ^{-1}(\tau t))\, , \qquad 
(\theta_2)_{\varphi,\psi}(\tau)=
\psi(\varphi ^{-1}(t+\tau (1- t)))\, , $$
$$(\theta_1)_{\varphi_1,\psi_1}(\tau)=\psi_1(\varphi_1 ^{-1}(\tau ))\, ,\qquad (\theta_2)_{\varphi_2,\psi_2}(\tau)=\psi_2(\varphi_2 ^{-1}(\tau ))\, .$$
\\
\\
{\bf Main theorem.}\\
For every bounded continuous function\\
$$F:C_+ ([0,1],\mathbb{R})\times C ([0,1],\mathbb{R})\times C_+ ([0,1],\mathbb{R})\times C ([0,1],\mathbb{R}) \to \mathbb{R}\, , \quad  \big( 0<t<1\,, \,\, \quad  a>0, \, b>0\big)\, ,$$
the following identity holds:
\begin{equation}\label{mainth}
    \int \limits _{0}^{+\infty}\!\int \limits _{\mathrm{Diff}_+^1([0,1])}\!\!
 \int \limits _{C([0,1],\mathbb{R})}\!\!
F((r_{1})_{\rho,\varphi},(\theta_{1})_{\varphi,\psi},
 (r_2)_{\rho,\varphi}, (\theta_{2})_{,\varphi,\psi}) \exp\left\{-\frac{a^2}{\sigma ^2}\rho^2\dot \varphi(0)-\frac{\sigma ^2 b}{8\rho^2}\right\}\times
\end{equation}
$$\rho\,\dot \varphi(0)\dot \varphi(1)\,\mu _{\frac{2\sigma}{\rho}}(d\varphi)\,
w_{\frac{\sigma}{\rho}}(d\psi)\, d\rho=$$
$$=\int \limits _{0}^{+\infty} \int \limits _{0}^{+\infty}\!
\int\limits _{\mathrm{Diff}_+^1([0,1])} \int\limits _{\mathrm{Diff}_+^1([0,1])}\!\!
\int \limits _{C([0,1],\mathbb{R})}\!\!\int \limits _{C([0,1],\mathbb{R})}\!\!
\delta\left(\rho_1\sqrt{\dot \varphi_1(1)}- \rho_2\sqrt{\dot \varphi_2(0)}\right)\times$$
$$\delta\left(\psi_1(1)- \psi_2(0)\right)
F( (r_1)_{\varphi _1,\rho_1},(\theta_1)_{\varphi_1,\psi_1},
 (r_2)_{\varphi _2,\rho_2}, (\theta_2)_{\varphi_2,\psi_2}) \exp\left\{-\frac{a^2}{t\sigma ^2}\rho_1^2\dot \varphi_1(0)-
\frac{tb\sigma ^2}{8\rho_1^2}-\frac{(1-t)b\sigma ^2}{8\rho_2^2}\right\}\times$$

$$\sqrt{\rho_1 \rho_2}\dot \varphi_1(0)(\dot \varphi_1(1)\dot \varphi_2(0))^{\frac{3}{4}}\dot \varphi_2(1)
\mu _{\frac{2\sigma \sqrt{t}}{\rho_1}}(d\varphi_1 )
\mu _{\frac{2\sigma \sqrt{1-t}}{\rho_2}}(d\varphi_2 )
w_{\frac{\sigma \sqrt{t}}{\rho_1}}(d\psi_1)
w_{\frac{\sigma \sqrt{1-t}}{\rho_2}}(d\psi_2) d\rho_1 d\rho_2.$$
\\

\newpage
\section{Fundamental solution of the heat equation on $\mathrm{Cone}$}\label{Fundamental solution} 
The theorem established above implies that the stochastic process $\xi(\tau)$ with path measure $\tilde w_{\sigma}$ has independent increments. Consequently, the measure $\tilde w_{\sigma}$ determines a one-parameter semigroup of operators of the form (\ref{U_t}) with the kernel (for simplicity, we set $\sigma=1$ below)
\begin{equation}\label{ConeK}
\tilde K(\tau,r,\theta,r_0,\theta_0)=\int_{C([0,\tau],\mathrm{Cone})} \, \delta^{(2)}(\xi(0)-y_{r_0,\theta_0})\delta^{(2)}(\xi(\tau)-x_{r,\theta})\,\,\tilde w_1(d\xi)\, ,\end{equation}
where 
\begin{equation}
   y_{r_0,\theta_0}=\left(
        r_0\cosh\theta_0\,,\,
        r_0\sinh\theta_0\right)\, ,\qquad x_{r,\theta}=\left(
        r\cosh\theta \, ,\,
        r\sinh\theta\right)\, .
\end{equation}
The kernel $\tilde K(\tau,r,\theta,r_0,\theta_0)$ is the fundamental solution of the heat equation 
\begin{equation}\label{heateq}
    \frac{\partial \tilde K}{\partial \tau}=\frac12\left(\frac{\partial^2}{\partial r^2}+\frac{1}{r}\frac{\partial}{\partial r}+\frac{1}{r^2}\frac{\partial^2}{\partial\theta^2}\right)\tilde K
\end{equation}
with the initial condition
\begin{equation}
   \tilde K(0,r,\theta,r_0,\theta_0)=\frac{1}{r}\delta(r-r_0)\delta(\theta-\theta_0)\, .
\end{equation}
Since the metric (\ref{Conemet1}) in the pseudopolar coordinates (\ref{x0x1hyp}) has the same form as the Euclidean metric on the plane (\ref{polarmet}), the operator on the right-hand side of (\ref{heateq}) is the Laplacian in polar coordinates. The pseudopolar angle $\theta\in \mathbb{R}$ ranges over the entire real line; hence no periodicity condition in $\theta$ is imposed on $\tilde K(\tau,r,\theta,r_0,\theta_0)$.  

We now reduce the integral (\ref{ConeK}) to a one-dimensional Riemann integral; see (\ref{Kint}) below. We pass from integration over $\xi\in C([0,\tau],\mathrm{Cone})$ to integration over $\eta\in C([0,1],\mathrm{Cone})$ by setting
\begin{equation}
    \xi(\tau_1)=\sqrt{\tau}\eta\left(\frac{\tau_1}{\tau}\right)\, .
\end{equation}

We use the relation 
\begin{equation}
    \tilde w_1(d\xi)=\tau \tilde w_1(d\eta)
\end{equation}
and obtain
\begin{equation}\label{Keta}
     \tilde K(\tau,r,\theta,r_0,\theta_0)=\frac{1}{\tau}\int_{C([0,1],\mathrm{Cone})} \delta^{(2)}\left(\eta(0)-\frac{y}{\sqrt{\tau}}\right)\delta^{(2)}\left(\eta(1)-\frac{x}{\sqrt{\tau}}\right)\,\,\tilde w_1(d\eta)\, .
\end{equation}

We express the process $\eta(\tau_1)$ in the decomposition coordinates $(\rho,\psi,\varphi)$:
\begin{equation}
    \eta(\tau_1)=\left(
         \frac{\rho}{\sqrt{\dot \varphi^{-1}(\tau_1)}}\cosh\left(\psi(\varphi^{-1}(\tau_1))\right)\, ,\,
         \frac{\rho}{\sqrt{\dot \varphi^{-1}(\tau_1)}}\,\,\sinh\left(\psi(\varphi^{-1}(\tau_1))\right)\right)\, .
\end{equation}
The measure $\tilde w_1(d\eta)$ on $C([0,1],\mathrm{Cone})$ takes the form
\begin{equation}\label{etadecomp}
     \tilde w_1(d\eta)=\rho \dot \varphi(0)\dot \varphi(1)\mu_{\frac{2}{\rho}}(d\varphi)w_{\frac{1}{\rho}}(d\psi)d\rho\, .
\end{equation}

Elementary transformations of the $\delta$-functions yield
\begin{equation}\label{deltafun}
     \delta^{(2)}\left(\eta(0)-\frac{y}{\sqrt{\tau}}\right)  \delta^{(2)}\left(\eta(1)-\frac{x}{\sqrt{\tau}}\right)=
\end{equation}     
\begin{equation*}     
     =\frac{4}{\rho^4}\delta\left(\dot\varphi(0)-\frac{r_0^2}{\tau\rho^2}\right)\delta\left(\dot \varphi(1)-\frac{r^2}{\tau\rho^2}\right)\delta(\psi(0)-\theta_0)\delta\left(\psi(1)-\theta\right)
\end{equation*}
and therefore allow us, in integral (\ref{Keta}), to make the substitution 
\begin{equation}\label{rhovarphi}
    \rho\dot \varphi(0)\dot \varphi(1)=\frac{r^2_0r^2}{\tau^2\rho^3}\, .
\end{equation}

Substituting (\ref{etadecomp}), (\ref{deltafun}), and (\ref{rhovarphi}) into (\ref{Keta}), we obtain

\begin{equation}\label{Kmuw}
   \tilde K(\tau,r,\theta,r_0,\theta_0)=\int d\rho \, \frac{4r^2r_0^2}{\tau^3\rho^7}\left\{\int\mu_{\frac{2}{\rho}}(d\varphi)\,\, \delta\left(\dot \varphi(0)-\frac{r_0^2}{\tau\rho^2}\right)\delta\left(\dot \varphi(1)-\frac{r^2}{\tau\rho^2}\right) \right\} \times
\end{equation}
\begin{equation*}
     \left[\int w_{\frac{1}{\rho}}(d\psi)\,\delta(\psi(0)-\theta_0)\delta(\psi(1)-\theta)\right]\, .
\end{equation*}

The integral with respect to the Wiener measure in square brackets is
\begin{equation}
    \int w_{\frac{1}{\rho}}(d\psi)\,\delta(\psi(0)-\theta_0)\delta(\psi(1)-\theta)=\frac{\rho}{\sqrt{2\pi}}\exp\left(-\frac{\rho^2(\theta-\theta_0)^2}{2}\right)\, .
\end{equation}
The integral over the diffeomorphism group in braces was evaluated in \cite{CorrSchwartz}:
\begin{equation}
    \mathcal{E}_{\sigma}(u,v)\equiv\int_{\mathrm{Diff}_{+}^1([0,1])}\delta(\dot\varphi(0)-u)\delta(\dot\varphi(1)-v)\mu_{\sigma}(d\varphi)=
\end{equation}
\begin{equation}
     ={\frac{8}{\sqrt{uv}(\sqrt{2\pi}\sigma)^3}}{\exp\left(-\frac{2}{\sigma^2}(u+v)\right)}\times
\end{equation}
\begin{equation}
  \int_0^{+\infty} d\chi \,\, {\exp\left(-\frac{4\sqrt{uv}}{\sigma^2}\cosh\chi\right)}\exp\left(\frac{2(\pi^2-\chi^2)}{\sigma^2}\right) \sin\left(\frac{4\pi \chi}{\sigma^2}\right)\sinh\chi\, .
\end{equation}

Thus, the integral in braces in (\ref{Kmuw}) is  
\begin{equation}
     \mathcal{E}_{2/\rho}\left(\frac{r_0^2}{\rho^2 \tau},\frac{r^2}{\rho^2 \tau}\right)={\frac{\tau\rho^5}{rr_0}\frac{1}{(2\pi)^{3/2}}}{\exp\left(-\frac{r^2+r_0^2}{2\tau}\right)}\times
\end{equation}
\begin{equation}
   \int_0^{+\infty} d\chi \,\,{\exp\left(-\frac{rr_0}{\tau}\cosh\chi \right)}\exp\left(\frac{\pi^2-\chi^2}{2}\rho^2\right)\sin(\pi\rho^2\chi)\sinh 
\chi\, .
\end{equation}
 The fundamental solution therefore takes the form 
\begin{equation}\label{Kansw}
    \tilde K(\tau,r,\theta,r_0,\theta_0)=\frac{4rr_0}{\tau^2}\frac{\exp\left(-\frac{r^2+r_0^2}{2\tau}\right)}{(2\pi)^2}\int_0^{+\infty}d\rho \,\,\frac{1}{\rho}\exp\left(-\frac{\rho^2}{2}(\theta-\theta_0)^2\right)\times
\end{equation}
\begin{equation*}
     \int_0^{+\infty}d \chi \exp\left(-\frac{rr_0}{\tau}\cosh\chi\right)\exp\left(\frac{\pi^2-\chi^2}{2}\rho^2\right)\sin(\pi\rho^2\chi)\sinh \chi\, .
\end{equation*}
In the integral over $\rho$, we make the substitution $\rho=\frac{1}{\sqrt{t}}$ to obtain 
\begin{equation}
    \tilde K(\tau,r,\theta,r_0,\theta_0)=\frac{\exp\left(-\frac{r^2+r_0^2}{2\tau}\right)}{2\pi \tau}\int_0^{+\infty} dt \frac{1}{\sqrt{t}} \exp\left(-\frac{(\theta-\theta_0)^2}{2t}\right) (2\pi)^{1/2}\Theta\left(\frac{rr_0}{\tau},t\right)\, ,
\end{equation}
where $\Theta\left(\frac{rr_0}{\tau},t\right)$ is the unnormalized Hartman--Watson density \cite{HW}, given by 
\begin{equation}
    \Theta(u,t)\equiv\frac{u}{\sqrt{2\pi^3t}}\int_0^{+\infty} d\chi \exp\left(-u\cosh\chi\right)\exp\left(\frac{\pi^2-\chi^2}{2t}\right)\sin\left(\frac{\pi\chi}{t}\right)\sinh\chi\, .
\end{equation}

The Laplace transform of $\Theta(u,t)$ is \cite{HW}, \cite{Yor}, \cite{HWLapl}
\begin{equation}
    \int_0^{+\infty} dt \exp\left(-\frac{t}{2}p^2\right)\Theta(u,t)=I_{|p|}(u)\, ,
\end{equation}
where $I_{p}(u)$ is the modified Bessel function of the first kind of order $p$.

Using the integral representation 
\begin{equation}
    \frac{1}{\sqrt{t}}\exp\left(-\frac{(\theta-\theta_0)^2}{2t}\right)=\frac{1}{\sqrt{2\pi}}\int_{-\infty}^{+\infty} dp \,\, \exp\left(-\frac{tp^2}{2}\right)\cos\left[p(\theta-\theta_0)\right]\, ,
\end{equation}
we obtain the following expression for the fundamental solution:
\begin{equation}\label{Kint}
    \tilde K(\tau,r,\theta,r_0,\theta_0)=\frac{\exp\left(-\frac{r^2+r_0^2}{2\tau}\right)}{2\pi\tau}\int_{-\infty}^{+\infty}dp\,\, \cos\left[p(\theta-\theta_0)\right]I_{|p|}\left(\frac{rr_0}{\tau}\right)\, . 
\end{equation}
Expression (\ref{Kint}) agrees with the results of \cite{Smits},\cite{Ford},\cite{DowQFTCone}.

The sum over all sheets of the covering space,
\begin{equation}\label{sumrots}
    \sum_{n\in\mathbb{Z}}\tilde K(\tau,r,\theta+2\pi n,r_0,\theta_0)=\frac{1}{2\pi\tau}\exp\left(-\frac{r^2+r_0^2-2rr_0\cos(\theta-\theta_0)}{2\tau}\right)
\end{equation}
coincides with the fundamental solution on $\mathbb{R}^2$ given in (\ref{2dkern}).

 It was shown in \cite{Comtet}--\cite{DowMulty} that a path integral on a multiply connected manifold $M$ can be expressed as a linear combination of integrals over homotopy classes $\mathcal{C}$ of paths in the universal cover $\tilde {M}$: 
\begin{equation}
    \int_{C([0,T],M)} F(\xi)\exp\left(-S(\xi)\right)d\xi=\sum_{\mathcal{C}}A(\mathcal{C})\int_{\xi\in \mathcal{C}} F(\xi)\exp\left(-S(\xi)\right)d\xi\, ,
\end{equation}
The coefficients $A(\mathcal{C})$ then define a representation of the fundamental group $\pi_1(M)$. The heat kernel is accordingly expressed as the sum 
\begin{equation}\label{KtildeKM}
    K(\tau,x,y)=\sum_{g\in\pi_1(M)}A(g)\tilde K(\tau,\tilde x,g\tilde y)\, ,
\end{equation}
where $\tilde x$ and $\tilde y$ are lifts of the corresponding points to the covering space. 

Identity (\ref{sumrots}) is fully consistent with this result. Since $\pi_1(\mathbb{R}^2\backslash\{0\})=\mathbb{Z}$, there is a family of heat kernels on $\mathbb{R}^2\backslash\{0\}$ of the form 
\begin{equation}\label{betakern}
    K_{\beta}(\tau,r,\theta,r_0,\theta_0)=\sum_{n\in \mathbb{Z}}e^{\beta n} \tilde K(\tau,r,\theta+2\pi n,r_0,\theta_0)\, .
\end{equation}
For $\beta=0$, we recover the familiar kernel (\ref{2dkern}). A detailed study of the properties of kernels of the form (\ref{betakern}) and the corresponding path measures $w^{(\beta)}_{\sigma}$ will be presented in future work.

\newpage

\section{Conclusion}\label{Conclusion}
In this work, we have constructed a measure on paths lying inside the cone: 
\begin{equation}
    \tilde w_{\sigma}(d\xi)=\rho\dot \varphi(0)\dot \varphi(T) \mu_{\frac{2\sigma}{\rho}}(d\varphi)w_{\frac{\sigma}{\rho}}(d\psi)d\rho=\exp\left(-\frac{1}{2\sigma^2}\int_0^Td\tau\,\, \langle \dot \xi, \dot \xi \rangle\right)d\xi\, ,
\end{equation}
where $\langle\dot \xi,\dot \xi \rangle$ is given by (\ref{Conemetric}). This measure is invariant under the Lorentz transformations (\ref{Lorentz}) and quasi-invariant under the action of the diffeomorphism group (\ref{pseudo act}). We have stated and proved Theorem (\ref{mainth}), which establishes that this measure defines a stochastic process with independent increments (a Markov process).

We have obtained the fundamental solution of the heat equation on $\mathrm{Cone}$: 
\begin{equation*}
    \tilde K(\tau,r,\theta,r_0,\theta_0)=\frac{\exp\left(-\frac{r^2+r_0^2}{2\tau}\right)}{2\pi\tau}\int_{-\infty}^{+\infty}dp\,\cos\left[p(\theta-\theta_0)\right] I_{|p|}\left(\frac{rr_0}{\tau}\right)\, .
\end{equation*}

We have shown that the space $\mathrm{Cone}$ equipped with the metric (\ref{Conemetric}) admits a bijective isometry onto the universal cover $\mathrm{Cover}$ of the punctured plane. We have described the geodesics on these spaces; paths near these geodesics make the dominant contribution to the path integral as $\sigma\to 0$.

Heat kernels are widely used in quantum field theory \cite{Ramond,Vassilevich} to compute effective actions, one-loop divergences, and quantum anomalies. Heat kernels on Riemann surfaces, of which the universal cover of the punctured plane is a particular example, arise naturally in quantum field theories on manifolds with singularities and in the presence of topologically nontrivial field configurations \cite{DowQFTCone}, \cite{DowThermalprop}--\cite{Furs}. In theories containing fermions, heat kernels can be used to compute the index of the Dirac operator \cite{Fri}, \cite{Berline}; a nonzero index indicates the presence of anomalies. Euclidean path integrals on Riemann surfaces arise naturally in connection with the replica approach to quantum field theory at finite temperature \cite{Callan}--\cite{Mald}. In future work, we will use the constructed measure $\tilde w_{\sigma}$ in some of these problems.

\vspace{1cm}
The research has been performed in the framework of the state assignment of M.V. Lomonosov
Moscow State University. The work of V. Chistiakov is supported by the Basis Foundation.

\clearpage
\begingroup
\setlength{\abovedisplayskip}{5pt plus 2pt minus 2pt}
\setlength{\belowdisplayskip}{5pt plus 2pt minus 2pt}
\setlength{\abovedisplayshortskip}{3pt plus 2pt}
\setlength{\belowdisplayshortskip}{3pt plus 2pt minus 1pt}
\section*{Appendix A}
To show that (\ref{wtilde}) and (\ref{tildewxi}) are equivalent, we use expression (\ref{mu}) for the quasi-invariant measure on the diffeomorphism group and expression (\ref{win1d}) for the Wiener measure. We write $\tilde w_{\sigma}$ as
\begin{equation}
     \tilde w_{\sigma}(d\xi)= \rho\sqrt{\dot \varphi(0)\dot \varphi(T)} \times
\end{equation}
\begin{equation*}
   \exp\left(\frac{\rho^2}{4\sigma^2}\left[\frac{\ddot \varphi(0)}{\dot\varphi(0)}-\frac{\ddot \varphi (T)}{\dot \varphi(T)}\right]+\frac{\rho^2}{4\sigma^2}\int_0^Tds\left\{\frac{d}{ds}\left(\frac{\ddot \varphi(s)}{\dot \varphi(s)}\right)-\frac12\left(\frac{\ddot \varphi(s)}{\dot \varphi(s)}\right)^2\right\}\right)\times
\end{equation*}
\begin{equation*}
\exp\left(-\frac{\rho^2}{2\sigma^2}\int_0^Tds\, (\dot \psi(s))^2\right)d\varphi\, d\psi\, d\rho\, .
\end{equation*}
Simplifying the exponent, we obtain
\begin{equation}\label{tildesmp}
    =\rho\sqrt{\dot \varphi(0)\dot \varphi(T)}\exp\left(-\frac{1}{2\sigma^2}\int_0^Tds\left\{\frac{\rho^2}{4}\left(\frac{\ddot \varphi(s)}{\dot \varphi(s)}\right)^2+\rho^2(\dot \psi(s))^2\right\}\right)\, .
\end{equation}
We change the integration variable from $s$ to $\tau$ by setting
\begin{equation}
s=\varphi^{-1}(\tau)\, .
\end{equation}
The derivatives satisfy
\begin{equation}
    \frac{d}{ds}\varphi(s)=\frac{1}{\dot \varphi^{-1}(\tau)}\, ,\qquad \frac{d^2}{ds^2} \varphi(s)=\frac{1}{\dot \varphi^{-1}(\tau)}\frac{d}{d\tau}\left(\frac{1}{\dot \varphi^{-1}(\tau)}\right)=-\frac{\ddot \varphi^{-1}(\tau)}{(\dot \varphi^{-1}(\tau))^3}\, .
\end{equation}
The integral in (\ref{tildesmp}) therefore satisfies
\begin{equation}\label{intint}
    \int_0^Tds\left\{\frac{\rho^2}{4}\left(\frac{\ddot \varphi(s)}{\dot \varphi(s)}\right)^2+\rho^2(\dot \psi(s))^2\right\}=\int_0^T d\tau \left\{\frac{\rho^2}{4}\frac{(\ddot \varphi^{-1}(\tau))^2}{(\dot \varphi^{-1}(\tau))^3}+\rho^2\dot \varphi^{-1}(\tau)(\dot \psi(\varphi^{-1}(\tau)))^2\right\}
\end{equation}
We now use (\ref{decompcone}) to express $\rho,\varphi$ in terms of $\xi^0,\xi^1$:
\begin{equation}
    \dot \varphi^{-1}(\tau)=\rho^2\frac{1}{(\xi^0(\tau))^2-(\xi^1(\tau))^2}\, ,\qquad \ddot \varphi^{-1}(\tau)=-2\rho^2\frac{\xi^0(\tau)\dot \xi^0(\tau)-\xi^1(\tau)\dot \xi^1(\tau)}{((\xi^0(\tau))^2-(\xi^1(\tau))^2)^2}\, .
\end{equation}
For the first term on the right-hand side of (\ref{intint}), we have
\begin{equation}\label{dr2}
    \frac{\rho^2}{4}\frac{(\ddot \varphi^{-1}(\tau))^2}{(\dot \varphi^{-1}(\tau))^3}=\frac{(\xi^0(\tau)\dot \xi^0(\tau)-\xi^1(\tau)\dot \xi^1(\tau))^2}{(\xi^0(\tau))^2-(\xi^1(\tau))^2}\, .
\end{equation}
Expression (\ref{decompcone}) for $\psi(s)$ gives
\begin{equation}
    \dot \psi(s)=\frac{d}{ds}\left(\mathrm{arctanh}\left(\frac{\xi^1(\varphi(s))}{\xi^0(\varphi(s))}\right)\right)=\frac{\dot \xi^1(\varphi(s))\xi^0(\varphi(s))-\dot \xi^0(\varphi(s))\xi^1(\varphi(s))}{(\xi^0(\varphi(s)))^2-(\xi^1(\varphi(s)))^2}\dot \varphi(s)\, ,
\end{equation}
\begin{equation}
    \dot \psi(\varphi^{-1}(\tau))=\frac{\dot \xi^1(\tau)\xi^0(\tau)-\dot \xi^0(\tau) \xi^1(\tau)}{(\xi^0(\tau))^2-(\xi^1(\tau))^2}\frac{1}{\dot \varphi^{-1}(\tau)}\, .
\end{equation}
The second term in (\ref{intint}) becomes
\begin{equation}\label{r2dpsi2}
    \rho^2\dot \varphi^{-1}(\tau)(\dot \psi(\varphi^{-1}(\tau)))^2=\frac{(\dot \xi^1(\tau)\xi^0(\tau)-\dot \xi^0(\tau) \xi^1(\tau))^2}{(\xi^0(\tau))^2-(\xi^1(\tau))^2}\, .
\end{equation}
Combining (\ref{dr2}) and (\ref{r2dpsi2}), we obtain
\begin{equation}
    \int_0^T d\tau \left\{\frac{\rho^2}{4}\frac{(\ddot \varphi^{-1}(\tau))^2}{(\dot \varphi^{-1}(\tau))^3}+\rho^2\dot \varphi^{-1}(\tau)(\dot \psi(\varphi^{-1}(\tau)))^2\right\}=
\end{equation}
\begin{equation*}
    =\int_0^Td\tau\left\{\frac{(\xi^0(\tau)\dot \xi^0(\tau)-\xi^1(\tau)\dot \xi^1(\tau))^2}{(\xi^0(\tau))^2-(\xi^1(\tau))^2}+\frac{(\dot \xi^1(\tau)\xi^0(\tau)-\dot \xi^0(\tau) \xi^1(\tau))^2}{(\xi^0(\tau))^2-(\xi^1(\tau))^2}\right\}=
\end{equation*}
\begin{equation*}
=\int_0^T d\tau \left\{(\dot \xi^0)^2-(\dot \xi^1)^2+2\frac{(\xi^0\dot \xi^1-\xi^1\dot \xi^0)^2}{(\xi^0)^2-(\xi^1)^2}\right\}\, .
\end{equation*}

\endgroup
\clearpage
\begingroup
\footnotesize
\setlength{\abovedisplayskip}{3pt plus 1pt minus 1pt}
\setlength{\belowdisplayskip}{3pt plus 1pt minus 1pt}
\setlength{\abovedisplayshortskip}{2pt plus 1pt}
\setlength{\belowdisplayshortskip}{2pt plus 1pt minus 1pt}
\section*{Appendix B. Proof of the main theorem}

$$U( \varphi_1,\rho_1,
 \varphi _2,\rho_2)
= \int \limits _{C([0,1],\mathbb{R})}\!\!\int \limits _{C([0,1],\mathbb{R})}\!\!
\delta\left(\psi_1(1)- \psi_2(0)\right)
F( (r_1)_{\varphi _1,\rho_1},(\theta_1)_{\varphi_1,\psi_1},
 (r_2)_{\varphi _2,\rho_2}, (\theta_2)_{\varphi_2,\psi_2})\times$$
$$\exp\left\{-
\frac{t(b-1)\sigma ^2}{8\rho_1^2}-\frac{(1-t)(b-1)\sigma ^2}{8\rho_2^2}\right\} \sqrt{\rho_1 \rho_2}(\dot \varphi_1(0)\dot \varphi_2(1))^{\frac{1}{4}}
w_{\frac{\sigma \sqrt{t}}{\rho_1}}(d\psi_1)
w_{\frac{\sigma \sqrt{1-t}}{\rho_2}}(d\psi_2) =$$
$$
=
\int \limits _{C([0,1],\mathbb{R})}\!\!
F( (r_1)_{\varphi _1,\rho_1},(\theta_1)_{\varphi_1,\psi_1},
 (r_2)_{\varphi _2,\rho_2}, (\theta_2)_{\varphi_2,\psi_2})\times $$
$$\exp\left\{-
\frac{t(b-1)\sigma ^2}{8\rho_1^2}-\frac{(1-t)(b-1)\sigma ^2}{8\rho_2^2}\right\}\,\sqrt{\rho_1 \rho_2}\,(\dot \varphi_1(0)\dot \varphi_2(1))^{\frac{1}{4}}\,
w_{\frac{\sigma }{\rho}}(d\psi).
$$
Here 
$$\rho=\frac {\rho_1 \rho_2}{\sqrt{t\rho_2^2+(1-t)\rho_1^2}},$$
$$\psi_{1}(\tau)=\psi\left(\frac{t\rho^2}{\rho_1^2}\tau \right)\, ,\quad  \psi_{2}(\tau)=
\psi\left(\frac{t\rho^2}{\rho_1^2}+\frac{(1-t)\rho^2}{\rho_2^2}\tau\right)\, .$$\\
Then
$$U( \varphi _{1,\varphi,\rho},\rho_{1,\varphi,\rho},
 \varphi _{2,\varphi,\rho},\rho_{2,\varphi,\rho}) =
 \int \limits _{C([0,1],\mathbb{R})}\!\!
F((r_{1})_{\rho,\varphi},(\theta_{1})_{\varphi,\psi},
 (r_2)_{\rho,\varphi}, (\theta_2)_{\varphi,\psi})\times$$
$$\exp\left\{-\frac{\sigma ^2 (b-1)}{8\rho^2}\right\}
\rho(\dot \varphi(0)\dot \varphi(1))^{\frac{1}{4}}
w_{\frac{\sigma}{\rho}}(d\psi)\, , $$

$$I=\int \limits _{0}^{+\infty} \int \limits _{0}^{+\infty}\!
\int\limits _{\mathrm{Diff}_+^1([0,1])} \int\limits _{\mathrm{Diff}_+^1([0,1])}\!\!
\int \limits _{C([0,1],\mathbb{R})}\!\!\int \limits _{C([0,1],\mathbb{R})}\!\!
\delta\left(\rho_1\sqrt{\dot \varphi_1(1)}- \rho_2\sqrt{\dot \varphi_2(0)}\right)\times$$
$$\delta\left(\psi_1(1)- \psi_2(0)\right)
F( (r_1)_{\varphi _1,\rho_1},(\theta_1)_{\varphi_1,\psi_1},
 (r_2)_{\varphi _2,\rho_2}, (\theta_2)_{\varphi_2,\psi_2}) \exp\left\{-\frac{a^2}{t\sigma ^2}\rho_1^2\dot \varphi_1(0)-
\frac{tb\sigma ^2}{8\rho_1^2}-\frac{(1-t)b\sigma ^2}{8\rho_2^2}\right\} \times $$
$$\sqrt{\rho_1 \rho_2}\dot \varphi_1(0)(\dot \varphi_1(1)\dot \varphi_2(0))^{\frac{3}{4}}\dot \varphi_2(1)
\mu _{\frac{2\sigma \sqrt{t}}{\rho_1}}(d\varphi_1 )
\mu _{\frac{2\sigma \sqrt{1-t}}{\rho_2}}(d\varphi_2 )
w_{\frac{\sigma \sqrt{t}}{\rho_1}}(d\psi_1)
w_{\frac{\sigma \sqrt{1-t}}{\rho_2}}(d\psi_2) d\rho_1 d\rho_2=$$
$$=\int \limits _{0}^{+\infty} \int \limits _{0}^{+\infty}\!
\int\limits _{\mathrm{Diff}_+^1([0,1])} \int\limits _{\mathrm{Diff}_+^1([0,1])}\!\!
\delta\left(\rho_1\sqrt{\dot \varphi_1(1)}- \rho_2\sqrt{\dot \varphi_2(0)}\right) U( \varphi _1,\rho_1,
 \varphi _2,\rho_2)\times$$
$$\exp\left\{-\frac{a^2}{t\sigma ^2}\rho_1^2\dot \varphi_1(0)-
\frac{t\sigma ^2}{8\rho_1^2}-\frac{(1-t)\sigma ^2}{8\rho_2^2}\right\} (\dot \varphi_1(0)\dot \varphi_1(1))^{\frac{3}{4}}(\dot \varphi_2(0)\dot \varphi_2(1))^{\frac{3}{4}}
\mu _{\frac{2\sigma \sqrt{t}}{\rho_1}}(d\varphi_1 )
\mu _{\frac{2\sigma \sqrt{1-t}}{\rho_2}}(d\varphi_2 )
 d\rho_1 d\rho_2. $$
The lemma in {\bf Section \ref{Maintheorem}} implies
\begin{equation*}
    I=\int \limits _{0}^{+\infty} \!
 \int\limits _{\mathrm{Diff}_+^1([0,1])}\!\!
U( \varphi _{1,\varphi,\rho},\rho_{1,\varphi,\rho},
 \varphi _{2,\varphi,\rho},\rho_{2,\varphi,\rho}) \exp\left\{-\frac{a^2}{\sigma ^2}\rho^2\dot \varphi(0)-
\frac{\sigma ^2}{8\rho^2}\right\} (\dot \varphi(0)\dot \varphi(1))^{\frac{3}{4}}
\mu _{\frac{2\sigma }{\rho}}(d\varphi )
 d\rho.
\end{equation*}
Hence,
$$I=\int \limits _{0}^{+\infty}\!\int \limits _{\mathrm{Diff}_+^1([0,1])}\!\!
 \int \limits _{C([0,1],\mathbb{R})}\!\!
F((r_1)_{\rho,\varphi},(\theta_1)_{\varphi,\psi},
 (r_2)_{\rho,\varphi}, (\theta_2)_{\varphi,\psi})\exp\left\{-\frac{a^2}{\sigma ^2}\rho^2\dot \varphi(0)-\frac{\sigma ^2 b}{8\rho^2}\right\}\times$$
$$ \rho\,\dot \varphi(0)\dot \varphi(1)\,\mu _{\frac{2\sigma}{\rho}}(d\varphi )
w_{\frac{\sigma}{\rho}}(d\psi)\, d\rho\, .$$

\endgroup
\clearpage
\section*{Appendix C}
\subsection*{Geodesics on the future cone}
In the semiclassical approximation ($\sigma\to 0$), the dominant contribution to integrals of the form (\ref{ConeK}) comes from geodesics of the metric (\ref{Conemet1}). We therefore consider the following problem. Let two points be given on $\mathrm{Cone}$: $A$ with Minkowski coordinates $(r_A,\theta_A)$ and $B$ with coordinates $(r_B,\theta_B)$. What is the form of the geodesic passing through these two points? The geodesics exhibit qualitatively different behavior depending on the angular separation $|\theta_A-\theta_B|$. We consider three cases.

\vspace{0.5cm}

{\bf Case 1: $|\theta_A-\theta_B|\le \pi$}\\
Since the metric (\ref{Conemet1}), expressed in the Minkowski coordinates (\ref{x0x1hyp}), takes the form (\ref{polarmet}) and coincides with the metric of the Euclidean plane, the geodesic equation in the coordinates $r,\theta$ is the same as the equation of a straight line in polar coordinates on the plane:
\begin{equation}\label{geodpl1}
    r\cos(\theta-\theta_0)=r_0\, .
\end{equation}
In this equation, $\theta$ takes values in
\begin{equation}
    \theta\in \left[-\frac{\pi}{2}+\theta_0,\frac{\pi}{2}+\theta_0\right]\, .
\end{equation}
It follows that two points $A,B\in \mathrm{Cone}$ can be joined by a smooth geodesic of the form (\ref{geodpl1}) only if their angular separation does not exceed $\pi$: $|\theta_A-\theta_B|\le\pi$.
Under the map from $\mathrm{Cone}$ to $\mathrm{Cover}$ constructed above, both points $A$ and $B$ lie on the same sheet of the cover, and the geodesic (\ref{geodpl1}) maps to a straight line (Fig.~2).

\begin{figure}[H]
    \centering
    \begin{subfigure}{0.45\textwidth}
        \centering
        \includegraphics[width=\linewidth]{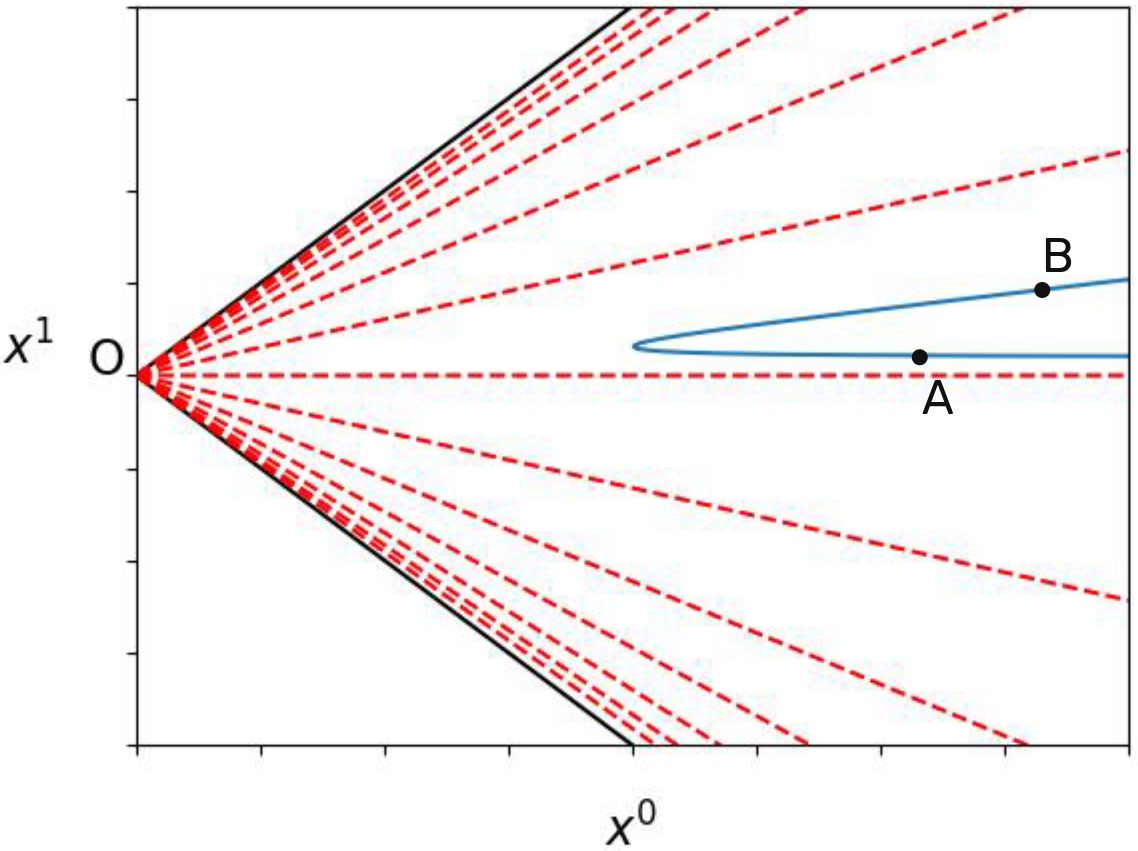}
        \caption{}
    \end{subfigure}
    \hfill
    \begin{subfigure}{0.45\textwidth}
        \centering
        \includegraphics[width=\linewidth]{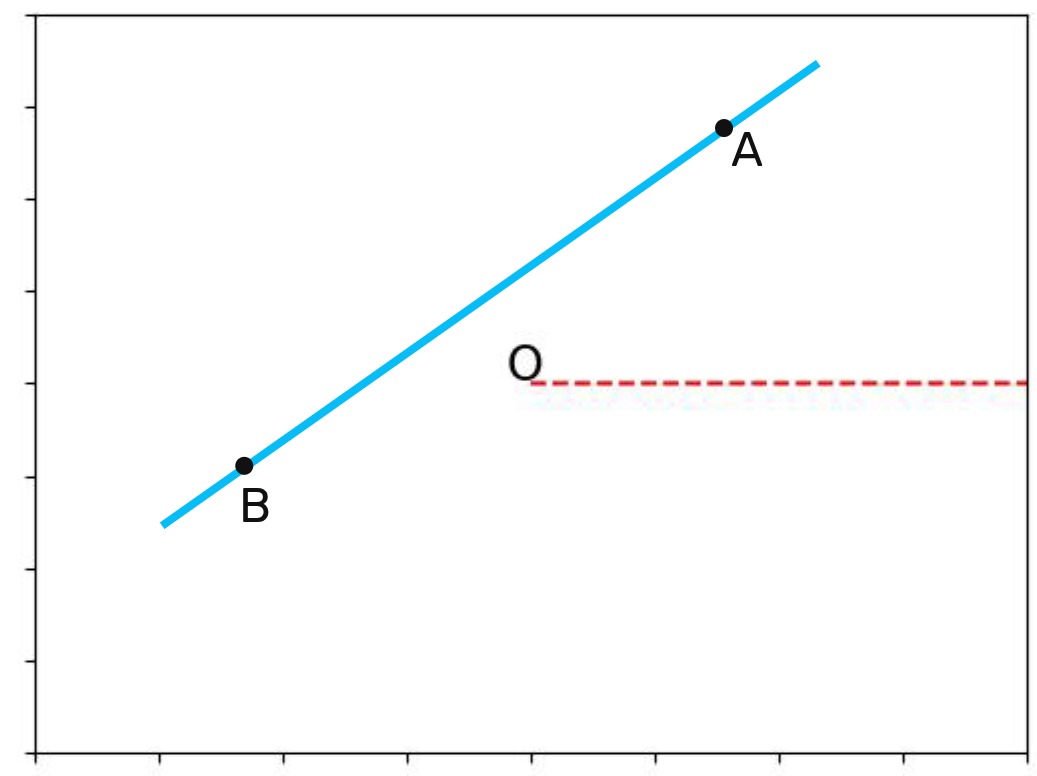}
        \caption{}
    \end{subfigure}
    \hfill
    \caption{The geodesic passing through points $A$ and $B$ when their angular separation is less than $\pi$. Panel (a) shows the geodesic on the cone, and panel (b) shows its image under the map to the universal cover of the punctured plane.}
    \label{fig:three_images_case1}
\end{figure}

\newpage

{\bf Case 2: $\pi<|\theta_A-\theta_B|<2\pi$}\\
It is convenient to consider both the points $A,B\in \mathrm{Cone}$ and their images in $\mathrm{Cover}$. For $|\theta_A-\theta_B|>\pi$, both points $A$ and $B$ lie on the same sheet of the cover. The geodesic on this sheet is readily constructed. Because the sheet has a cut, the straight line $AB$ is not a geodesic, since it would cross the cut. Any line segment that does not cross the cut has the shortest length among all curves joining its endpoints. Thus, the shortest curve joining $A$ and $B$ is a broken line consisting of two segments that join $A$ and $B$ to the origin $O$ (Fig.~3).

\begin{figure}[H]
    \centering
    \begin{subfigure}{0.45\textwidth}
        \centering
        \includegraphics[width=\linewidth]{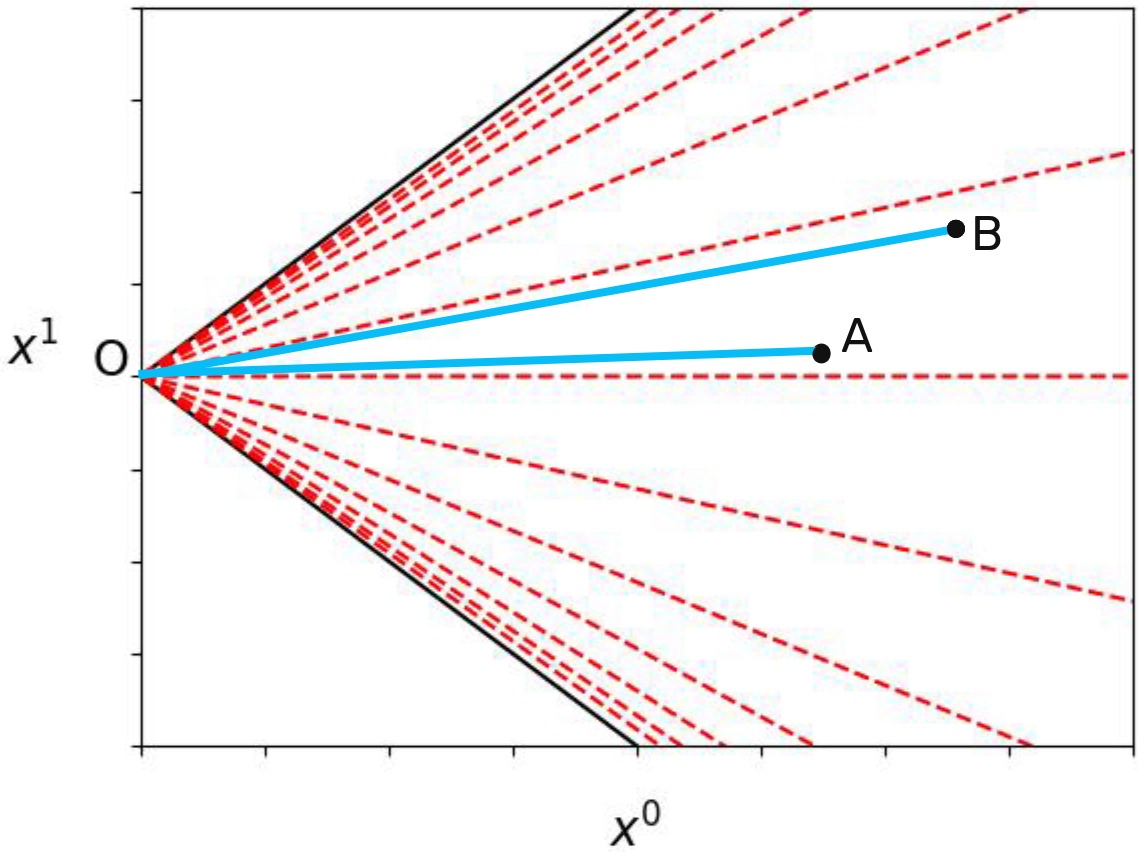}
        \caption{}
    \end{subfigure}
    \hfill
    \begin{subfigure}{0.45\textwidth}
        \centering
        \includegraphics[width=\linewidth]{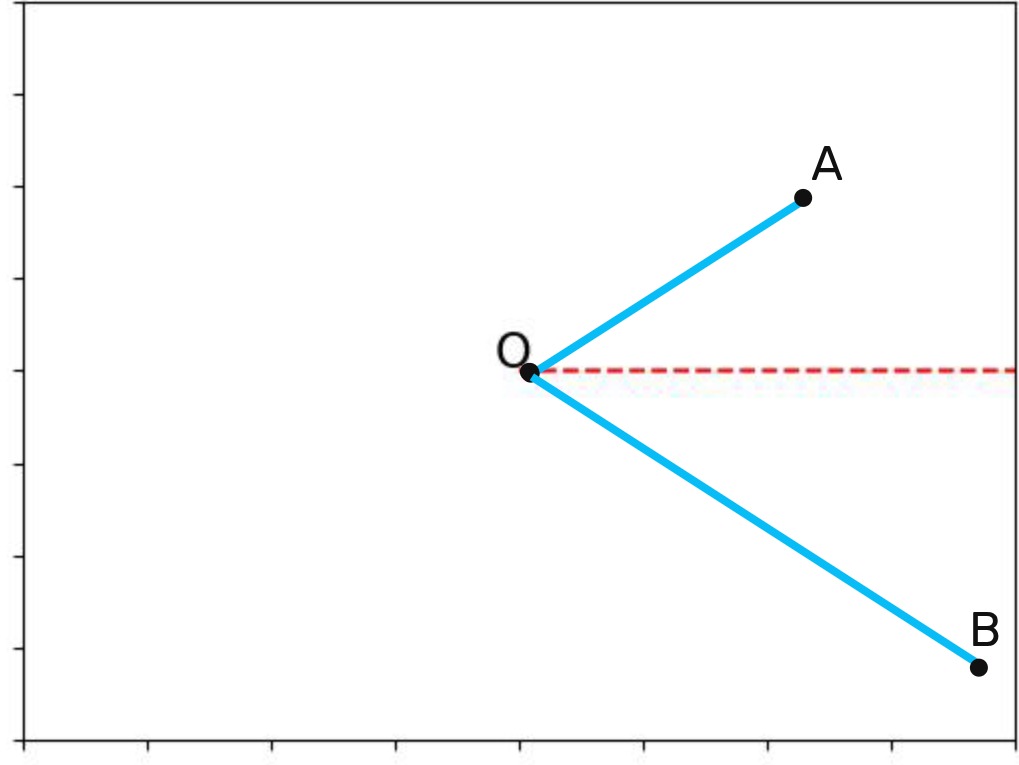}
        \caption{}
    \end{subfigure}
    \hfill
    \caption{The geodesic passing through points $A$ and $B$ when their angular separation is greater than $\pi$ but less than $2\pi$. Panel (a) shows the geodesic on the cone, and panel (b) shows its image under the map to the universal cover of the punctured plane. The images of $A$ and $B$ lie on the same sheet of the cover.}
    \label{fig:three_images_case2}
\end{figure}

\newpage

{\bf Case 3: $|\theta_A-\theta_B|>2\pi$} \\
In this case, the points $A,B$ lie on different sheets of the cover, and, as in the preceding case, the geodesic is a broken line consisting of two segments $OA$ and $OB$. Indeed, any other curve joining $A$ and $B$ (Fig.~4 (a)) is represented under the map to $\mathrm{Cover}$ by two arcs on different sheets of the cover that pass around the branch point $O$ (Fig.~4 (b), (c)). Clearly, $AO+OB<AM+MB$.

\begin{figure}[H]
    \centering
    \begin{subfigure}{0.45\textwidth}
        \centering
        \includegraphics[width=\linewidth]{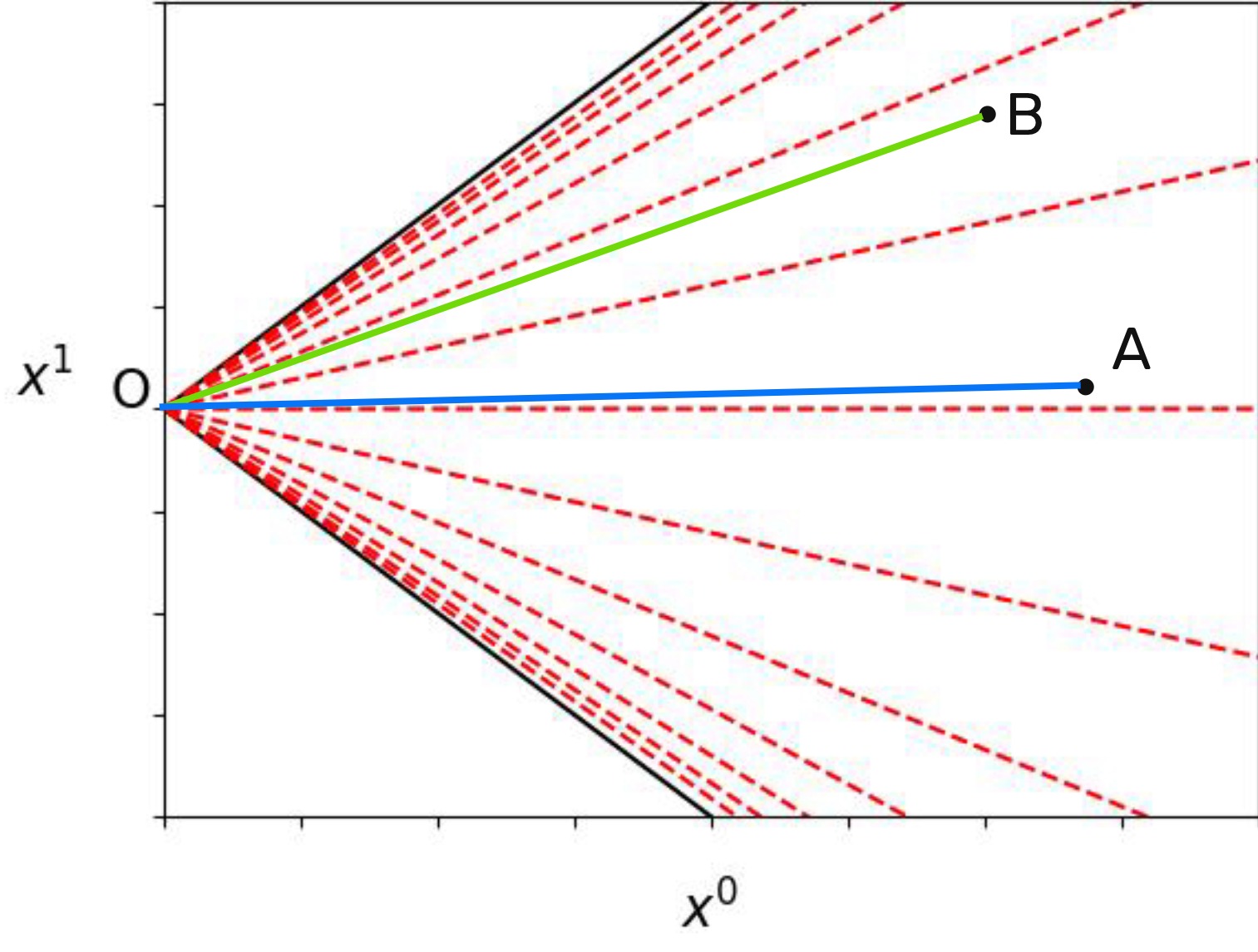}
        \caption{}
    \end{subfigure}
    \hfill
    \begin{subfigure}{0.45\textwidth}
        \centering
        \includegraphics[width=\linewidth]{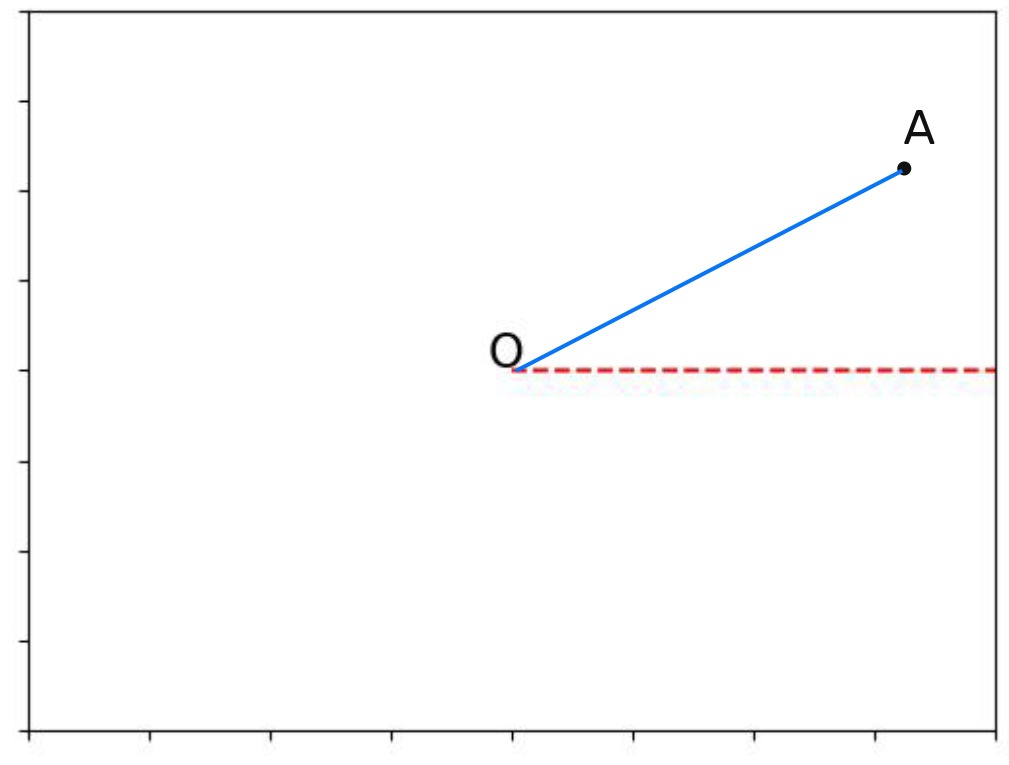}
        \caption{}
    \end{subfigure}
    \hfill
    \begin{subfigure}{0.45\textwidth}
        \centering
        \includegraphics[width=\linewidth]{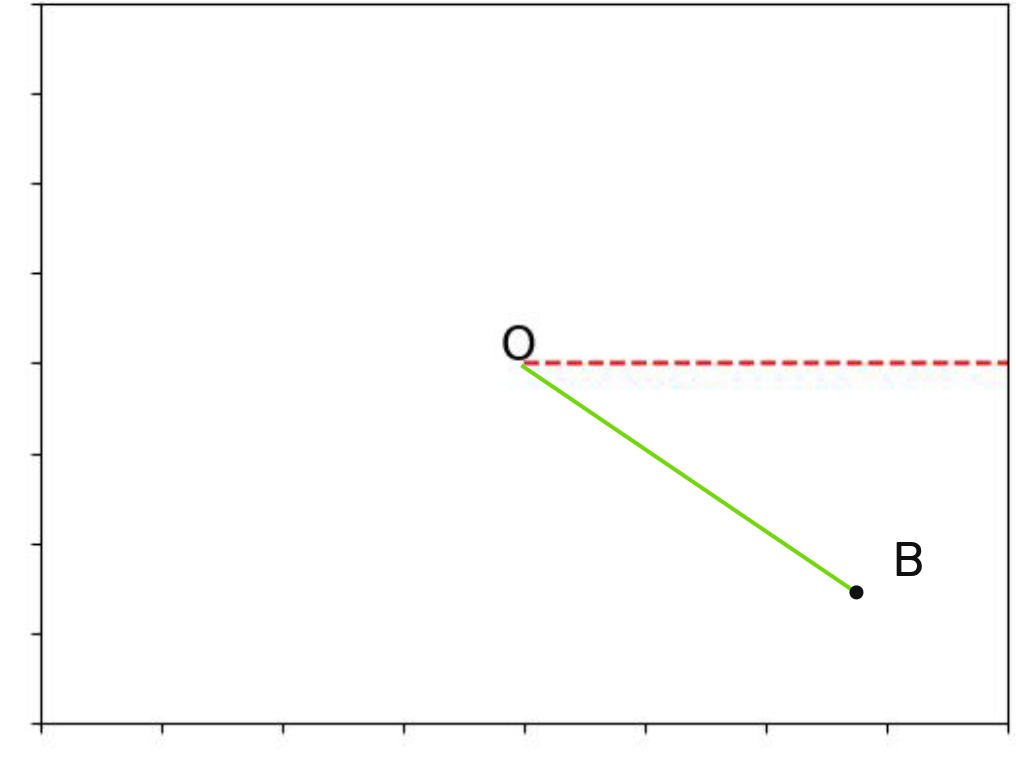}
        \caption{}
    \end{subfigure}
    \caption{The geodesic passing through points $A$ and $B$ when their angular separation is greater than $2\pi$. Panel (a) shows the geodesic on the cone, and panels (b) and (c) show its image under the map to the universal cover of the punctured plane. The images of $A$ and $B$ lie on different sheets of the cover.}
    \label{fig:three_images_case3}
\end{figure}

\begin{figure}[H]
    \centering
    \begin{subfigure}{0.45\textwidth}
        \centering
        \includegraphics[width=\linewidth]{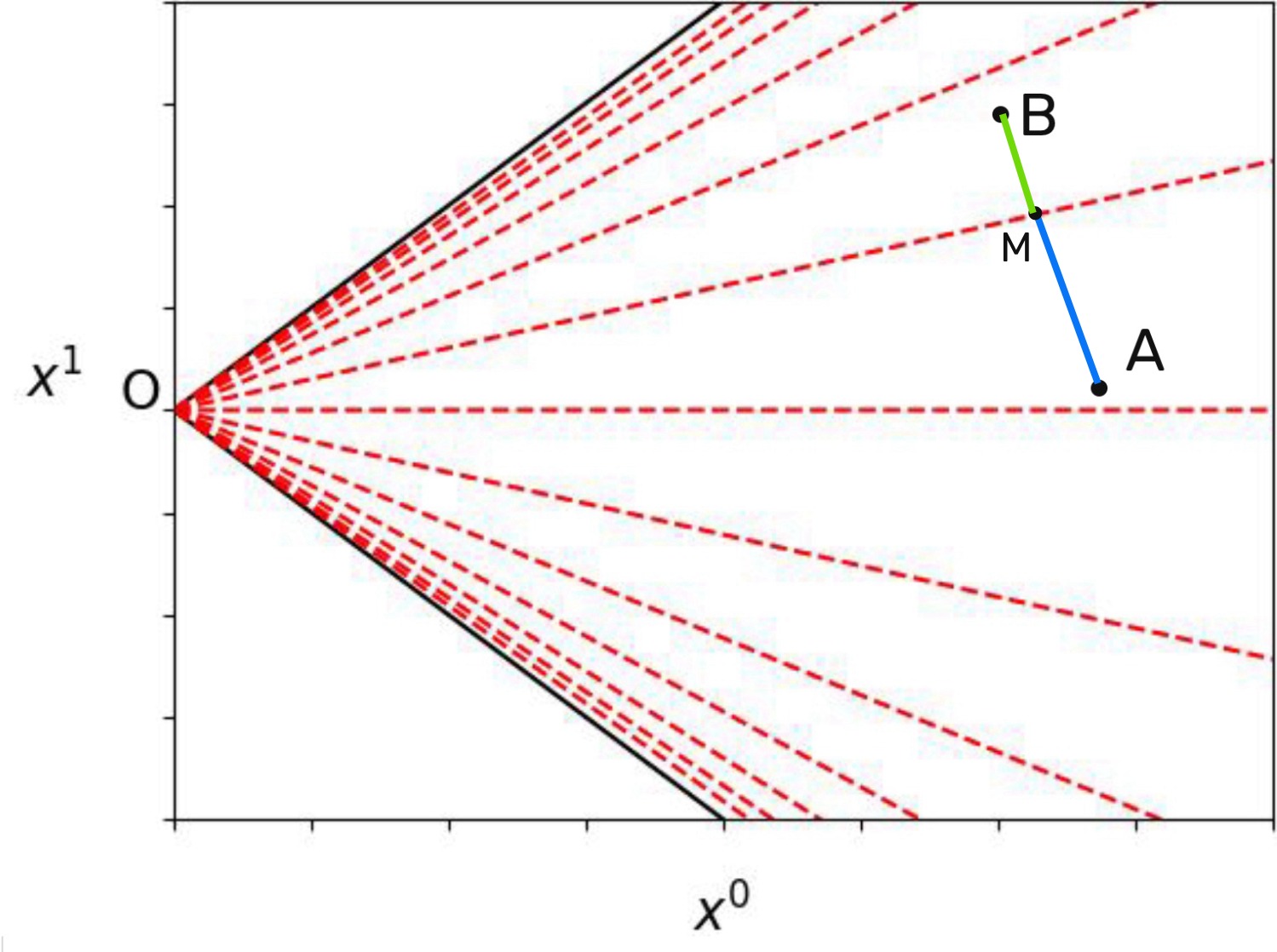}
        \caption{}
    \end{subfigure}
    \hfill
    \begin{subfigure}{0.43\textwidth}
        \centering
        \includegraphics[width=\linewidth]{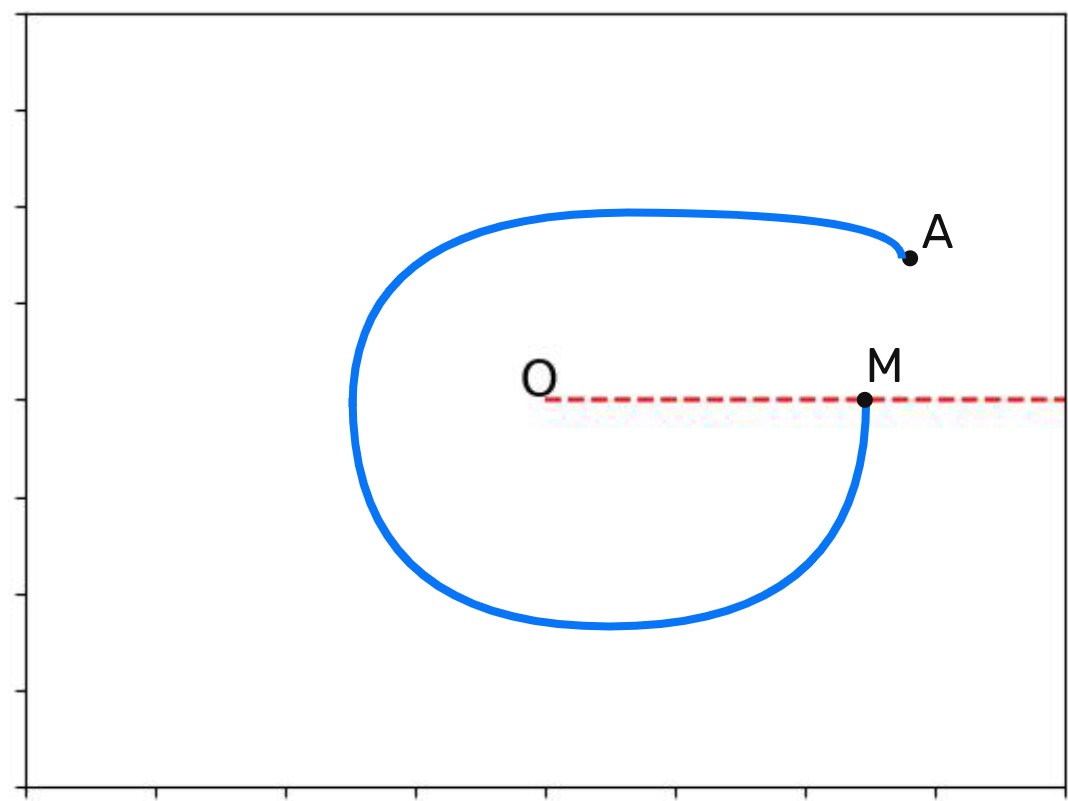}
        \caption{}
    \end{subfigure}
    \hfill
    \begin{subfigure}{0.43\textwidth}
        \centering
        \includegraphics[width=\linewidth]{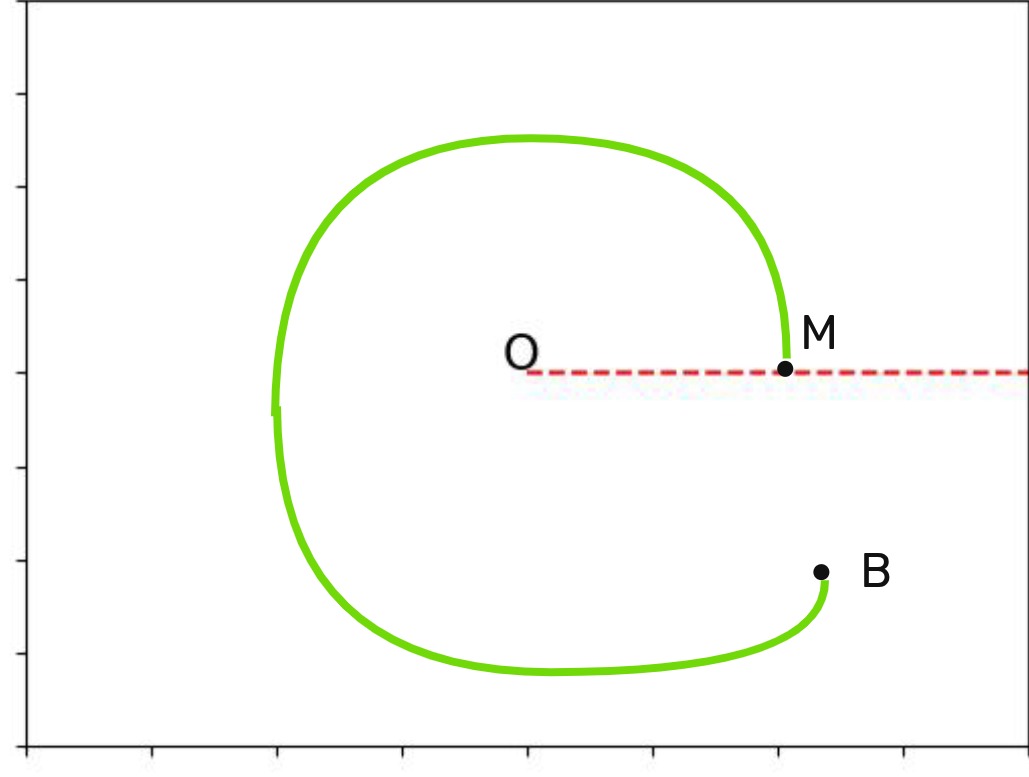}
        \caption{}
    \end{subfigure}
    \hfill
    \caption{An example of a curve joining points $A$ and $B$ with $|\theta_A-\theta_B|>2\pi$. Panel (a) shows the curve on the cone, and panels (b) and (c) show its image in the universal cover of the punctured plane. The images of $A$ and $B$ lie on different sheets of the cover.}
    \label{fig:three_images_other_curve}
\end{figure}

\newpage
\subsection*{Arbitrary paths on $\mathrm{Cone}$ and their images}
In the preceding subsection, we considered geodesics on the future cone. However, arbitrary paths also contribute to the path integrals. The correspondence (\ref{conecov}) allows us to regard the measure (\ref{wtilde}) constructed above as a measure on $C([0,T],\mathrm{Cover})$. We have thus obtained a tool for describing the motion of a quantum particle on the universal cover of the punctured plane. Integrals of the form (\ref{ConeK}), which describe the propagation of a particle from a point $A$ to a point $B$ on the same sheet of the cover, also receive contributions from paths that cross the cut and enter other sheets of the cover (Fig.~6).

\begin{figure}[H]
    \centering
    \begin{subfigure}{0.45\textwidth}
        \centering
        \includegraphics[width=\linewidth]{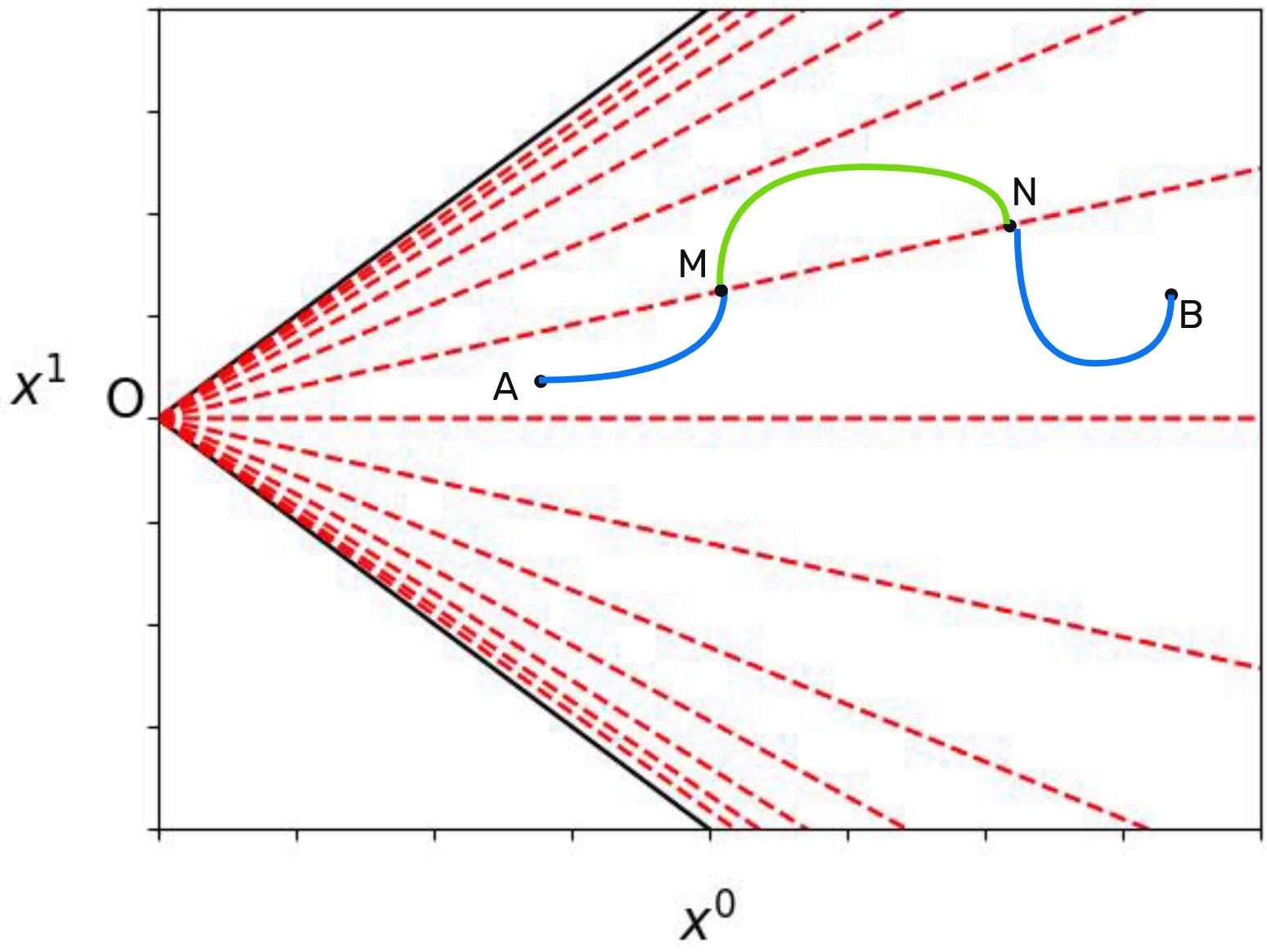}
        \caption{}
    \end{subfigure}
    \hfill
    \begin{subfigure}{0.43\textwidth}
        \centering
        \includegraphics[width=\linewidth]{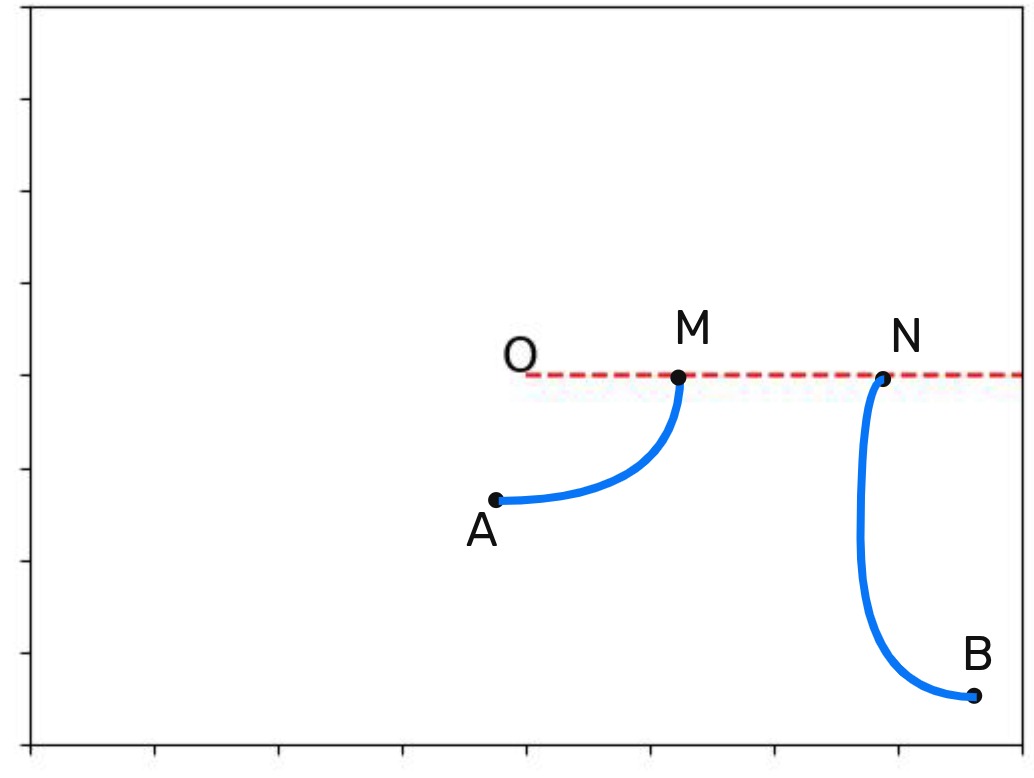}
        \caption{}
    \end{subfigure}
    \hfill
    \begin{subfigure}{0.43\textwidth}
        \centering
        \includegraphics[width=\linewidth]{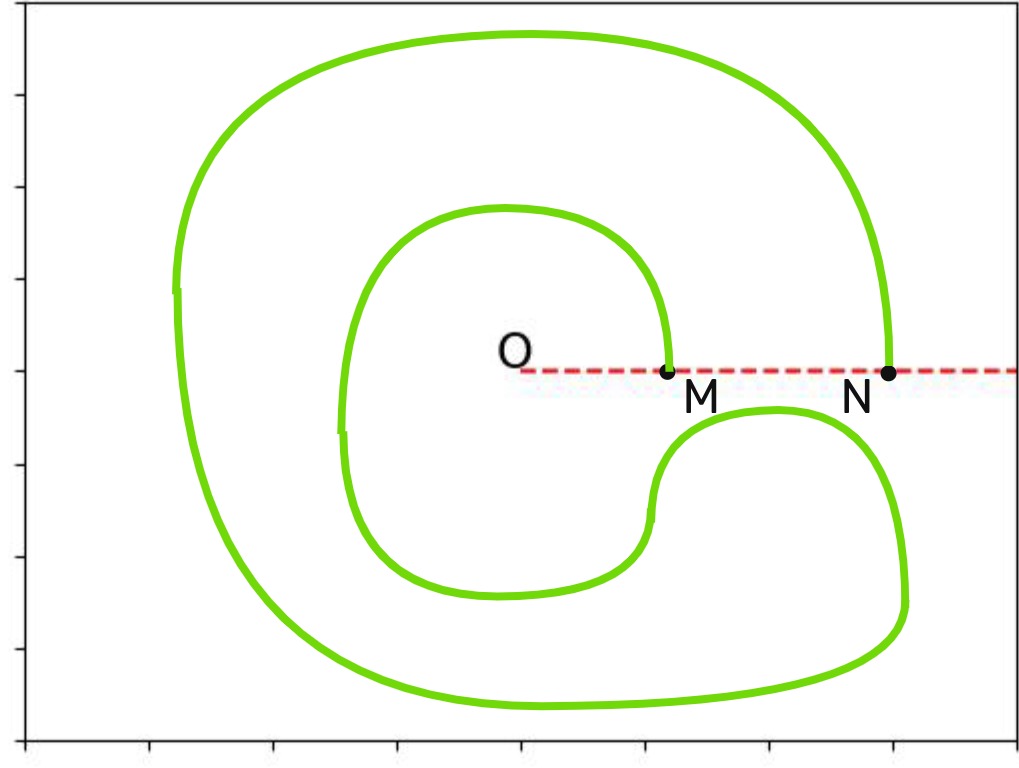}
        \caption{}
    \end{subfigure}
    \hfill
    \caption{An example of a path that joins two points on the same sheet of the cover while passing through different sheets. Panel (a) shows the path on the cone, and panels (b) and (c) show its image in the cover.}
    \label{fig:three_images_trajectory}
\end{figure}

\end{document}